\documentclass[%
 reprint,
 amsmath,amssymb,
 aps,
prx,
longbibliography
]{revtex4-1}
\usepackage{amsmath}

\usepackage{graphicx}
\usepackage{subfigure}
\usepackage{booktabs}
\usepackage{float}
\usepackage{dcolumn}

\newcommand{\net}[1]{{\textsc{#1}}}
\newcommand{\argmax}{\mathop{\mathrm{argmax}}\limits}

\begin{document}


\title{Phase transitions and optimal algorithms for semi-supervised classifications on graphs: from belief propagation to graph convolution network}

\author{Pengfei Zhou}
\affiliation{
 CAS Key Laboratory for Theoretical Physics, Institute of Theoretical Physics, Chinese Academy of Sciences, Beijing 100190, China
}
\affiliation{
 School of Physical Sciences, University of Chinese Academy of Sciences, Beijing 100049, China
}
\author{%
Tianyi Li}
\affiliation{
 System Dynamics Group, Sloan School of Management, \\
 Massachusetts Institute of Technology, Cambridge, MA 02139, USA 
 }
\author{Pan Zhang}%
 \email{panzhang@itp.ac.cn}
\affiliation{
 CAS Key Laboratory for Theoretical Physics, Institute of Theoretical Physics, Chinese Academy of Sciences, Beijing 100190, China
}


%
%
\date{\today}

\begin{abstract}
We perform theoretical and algorithmic studies for the problem of clustering and semi-supervised classification on graphs with both pairwise relational information and single-point feature information, upon a joint stochastic block model for generating synthetic graphs with both edges and node features.
Asymptotically exact analysis based on the Bayesian inference of the underlying model are conducted, using the cavity method in statistical physics. Theoretically, we identify a phase transition of the generative model, which puts fundamental limits on the ability of all possible algorithms in the clustering task of the underlying model. Algorithmically, we propose a \textit{belief propagation} algorithm that is asymptotically optimal on the generative model, and can be further extended to a \textit{belief propagation graph convolution neural network} (BPGCN) for semi-supervised classification on graphs. 

For the first time, well-controlled benchmark datasets with asymptotially exact properties and optimal solutions could be produced for the evaluation of graph convolution neural networks, and for the theoretical understanding of their strengths and weaknesses. In particular, on these synthetic benchmark networks we observe that existing graph convolution neural networks are subject to an sparsity issue and an overfitting issue in practice, both of which are successfully overcome by our BPGCN. Moreover, when combined with classic neural network methods, BPGCN yields extraordinary classification performances on some real-world datasets that have never been achieved before.

\end{abstract}

\maketitle

\section{Introduction}
Learning on graphs is an important task in machine learning and data sciences with a lot of applications in various fields, including social sciences (e.g., social network analysis), biology (e.g., protein structure prediction and molecular finger prints learning), and computer science (e.g., knowledge graph analysis). The key difference between learning with graph data and conventional machine learning with images and natural languages is that, in addition to content features on each item, there are also relational features between items that are encoded by edges in the graph, which adds an extra layer of complexity to the analysis.

One classical problem of learning on graphs is the classification of nodes into groups. Consider such a problem in the setting of a citation network, where each article is represented by a graph node, and the groups of nodes are scientific research fields. In addition to the edges between nodes, which represent the citations between articles, each node is also associated with some \textit{features} (i.e., key words), which encode typical information of research fields. If the group information of a small subset of nodes are known, then practically, these nodes could serves as a training set, and the learning task is to determine the group membership of the rest of the nodes by exploring the direct group information via features, as well as indirect information via relationships (represented by edges of the graph) with the training nodes.
Essentially, this learning task is  \textit{semi-supervised classification on graphs}, a problem that recently has drawn much attention in both networks sciences and machine learning communities; on this problem, we witnessed the burst of \textit{Graph Convolution Neural Networks} (GCN), which is a powerful neural network architecture that yields ground-breaking performances \cite{kipf2017semi}.

Deep convolution neural networks have made great success in machine learning and artificial intelligence ~\cite{goodfellow2016deep}. Since there are many situations in which data are represented as graphs, rather than voices and images that on one or two-dimensional grids, a lot of efforts have been made to extend convolution networks from applying on grid data to applying on graph data, with a heavy focus on constructing linear convolution kernels to extract local features in graphs and on learning effective representations of graph objects. During the past several years, many GCNs have been proposed, using different types of convolution kernels and different network architectures \cite{klicpera2018combining,2018graph,kipf2017semi,gilmer2017neural}. Recent studies showed that GCNs have quickly dominated among neural network techniques in the performance on various learning tasks, including text or graph-object classification, link prediction, forecasting, importance sampling, and are also believed to have a big potential on relational reasoning~\cite{battaglia2018relational}. Nevertheless, despite that GCNs have achieved the state-of-the-art performance on semi-supervised classification, so far there is little theoretical understanding on the mathematical principles behind graph convolutions, and on the extent that they may probably work in a particular problem setting. The main difficulty is that, in previous studies, GCNs are often only tested on real-world datasets which do not have clear theoretical structures, thus the success or failure of GCNs is hard to be pinned down in theoretical analysis. A set of network datasets with 
established mathematical properties,
and the analysis of GCNs on such datasets are missing and greatly welcomed.
 
In this study, we propose to study  semi-supervised classification on graphs, based on the celebrated stochastic block model (SBM) \cite{holland1983stochastic}. The joint model consists of two graph components, characterizing both the relational information and the feature information of items, captured respectively by a standard SBM and a bi-partite SBM \cite{hric2016network, florescu2016spectral}. We call the model Joint Stochastic Block Model (JSBM). This model was originally proposed in~\cite{hric2016network} for the problem of link and node predictions through the Markov chain Monte Carlo method. In this study we analyze theoretical properties of the JSBM,  and design techniques for semi-supervised learning on graphs based on its properties. Filling the gap discussed above, the JSBM produces well-controlled benchmark graphs with continuously tunable parameters for the evaluation of GCN's classification performance, and for the theoretical understanding of their strengths and weaknesses under certain conditions.

On graphs generated by the JSBM, the clustering and classification problems can be translated to a Bayesian inference problem, which could be solved theoretically with the statistical physics approach in an asymptotically exact manner. This approach leads to a message-passing algorithm, known in computer science as the belief propagation (BP) algorithm \cite{yedidia2003understanding}, which we claim to be asymptotically exact on large random graphs generated by the JSBM. Through analyzing the stability of fixed points of the constructed BP equations on the JSBM, a phase transition --- the \textit{detectability transition} is identified; beyond the phase transition point, no algorithm is able to conduct successful clustering on JSBM in an unsupervised manner. This is an extension of the detectability phase transition~\cite{decelle2011asymptotic} in the standard SBM~\cite{holland1983stochastic}, and puts fundamental limits on the ability of algorithms in the clustering tasks on graphs that could be modeled by JSBM.

In the semi-supervised classification setting, where a small fraction of nodes have ground-true group labels and could be used as the training data, the BP algorithm for JSBM could be embedded into a graph convolution network architecture. The unknown generative parameters of the JSBM graph could be learned in a standard classification approach, through forward-passing of (truncated) BP equations together with backward-passing of the gradients of the loss function. This novel GCN algorithm, which we term as BPGCN, guarantees to yield Bayes optimal classification results ~\cite{Zhang2014phase} on synthetic graphs generated by the JSBM, and performs comparably with state-or-the-art GCNs on real-world networks.

The paper is organized as follows. In Sec.~\ref{sec:jsbm}  we introduce the joint stochastic block model. In Sec.~\ref{sec:infer} we formulate the Bayesian inference problem for clustering and classification on the joint stochastic block model, and derive the belief propagation equations for the JSBM. In Sec.~\ref{sec:transition} we study the detectability phase transition of JSBM using stability analysis of the BP algorithm. In Sec.~\ref{sec:bpgcn} we convert the BP equations on JSBM to a graph convolution neural network and propose a new GCN algorithm, BPGCN. In Sec.~\ref{sec:comparing} the performance of BPGCN is evaluated and compared with the performance of several state-of-the-art GCNs, on both synthetic and real-world networks. Sec.~\ref{sec:con} concludes the study.

\section{Joint stochastic block model}\label{sec:jsbm}
The idea of the JSBM, literally the joint of two stochastic block models,
 is to simultaneously model item nodes and feature nodes in the network setting, 
 by representing both a connectivity graph over item nodes and an attribute graph between item nodes and feature nodes. 
 The connectivity graph corresponds to the relation network in the traditional sense, 
 and the attribute graph is a bipartite graph established on top of the relation network;
 both graphs associate each node in the graph with a unique group membership.
 These two graphs, constructed from two SBM processes, constitute the JSBM graph $\mathcal{G}$; 
 a similar framework was proposed by \cite{hric2016network} to study link predictions. 

Assume $n$ item nodes in the (undirected) connectivity graph, 
belonging to $\kappa$ groups. Each node $i$ has an unknown label $t_i^*$ denoting its independent group
 membership (i.e. $t_i^* \in \{ 1,2,...\kappa \}$); each group label is chosen at random by nodes, 
 according to a $\kappa$-dimensional probability vector $\mathbf{\alpha}$, whose entries sum to 1. 
 Edge connectivities are exclusively determined by their group memberships: for each pair of nodes $i$ and $j$,
 there exists an edge $(i,j)\in \mathcal{E}$ with probability $p_{t_i^*t_j^*}$, 
 where $\mathcal{E}$ is the entire edge set of the graph. 
 This stochastic generative process could be understood as a measuring process for the ground truth $t_i^*$, 
 whose information is encoded implicitly in the edge connectivities as measuring results. 
 The $n \times n$ adjacency matrix of this connectivity graph $\mathbf{A}\in \{0,1\}^{n\times n}$ follows the 
 generative process and is therefore stochastic, controlled by the deterministic $\kappa \times \kappa$ generation
 probability matrix $\mathbf{P} = \{p_{t_i^*t_j^*}\}$. If the diagonal elements in $\mathbf{P}$ are larger than the off-diagonal elements,
 it corresponds to the situation (known as \textit{assortative} SBM) where there are more edges within node groups than between groups, 
 and vice versa (known as \textit{disassortative} SBM). An intuitive example for an \textit{assortative} SBM is a citation
 network with nodes denoting research articles (items), which belong to a certain research area (group) and are linked through
 citations (edges). Apparently, articles from the same research area are more likely to cite each other.

\begin{figure}
\centering
\includegraphics[width=0.4\textwidth]{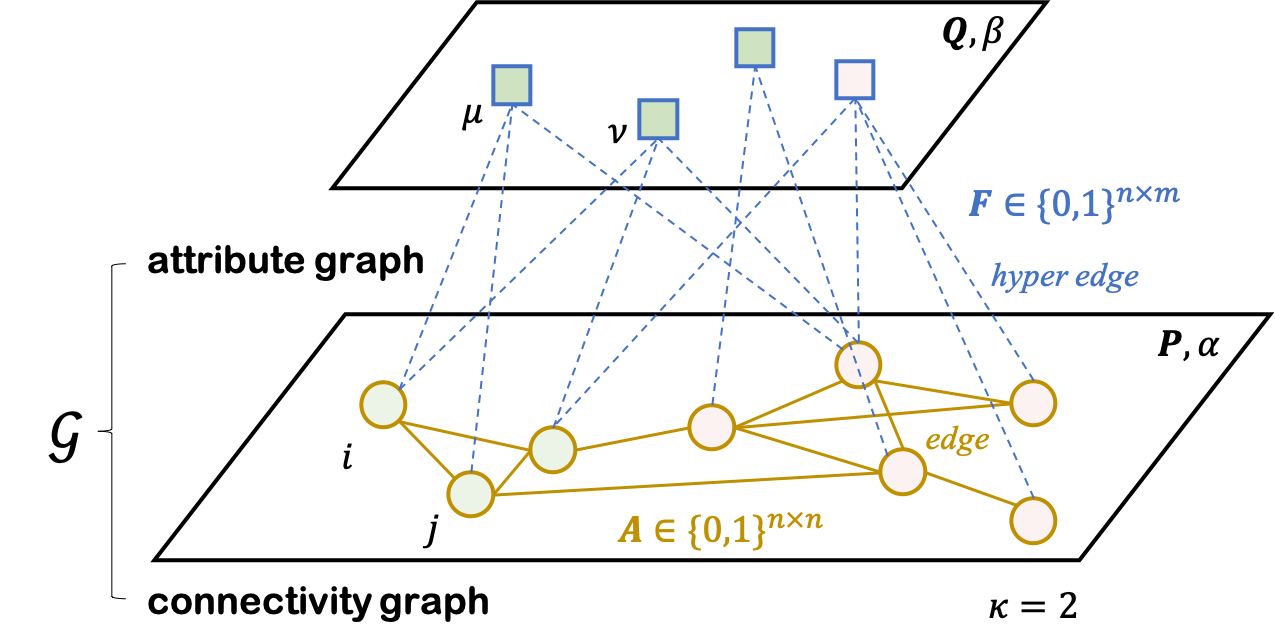}
\caption{Illustration of the joint stochastic block model (JSBM). Circles denote item nodes, whose inter-connections constitute the connectivity graph, with adjacency matrix $\mathbf{A}$ defined over edges; boxes denote feature nodes. The attribute graph is the bi-partite graph between items and features, with adjacency matrix $\mathbf{F}$ defined over \textit{hyper edges}. The JSBM graph is generated by parameters $\theta = \{\mathbf{P},\mathbf{Q}, \mathbf{\alpha}, \mathbf{\beta}\}$, with $\kappa$ groups of nodes built in for both items and features. \label{fig:fg}}
\end{figure}

Besides having membership in a certain research area, moreover, 
each research article may also be associated with some keywords which further denote its categorical information; 
expectedly, the research area that an article belongs to could be inferred from these keywords. 
This notion underlines the idea of making inference on the joint SBM, i.e. inferring hidden
 group membership of an item node from its relationships to known feature nodes in the attribute graph. 
 Assume such a graph with $m$ features over $n$ item nodes. Same as item nodes, 
 each feature node $\mu$ is embedded with an independent label $t_{\mu}^* \in \{ 1,2,...\kappa \}$, 
 chosen randomly from the $\kappa$ groups according to probabilities in a $\kappa$-dimensional vector $\mathbf{\beta}$. 
 Relationship between an item node $i$ and a feature node $\mu$ exists with probability $q_{t_i^*t_{\mu}^*}$, 
 which corresponds to an edge $(i,\mu)\in \mathcal{F}$ in the attribute graph whose edge set is $\mathcal{F}$. 
 This generative process yields a bipartite graph between item nodes and feature nodes, analogous to the bipartite SBM \cite{florescu2016spectral};
 the resulting $n \times m$ adjacency matrix of the attribute graph $\mathbf{F}\in \{0,1\}^{n\times m}$ is 
 stochastic and is governed by a $\kappa\times \kappa$ matrix $\mathbf{Q} = \{q_{t_i^{*} t_{\mu}^*}\}$, 
 similar to the case of $\mathbf{A}$ and $\mathbf{P}$. A specific JSBM is thus represented by
 the two adjacent matrices $\mathbf{A}$ and $\mathbf{F}$ for the connectivity graph and attribute graph, 
 respectively, and is controlled by the generation parameters $\theta = \{\mathbf{P},\mathbf{Q}, \mathbf{\alpha}, \mathbf{\beta}\}$ (Fig.~\ref{fig:fg}).
 It consists of a uni-partite graph and a bipartite graph, similar to the structure in semi-restricted Boltzmann machines \cite{osindero2008modeling}. 
 In this setting, a feature $\mu$ plays the role of a hidden variable, or a functional node collecting multiple
 interactions with the item nodes connected to it; hence edges on 
 the attribute graph, i.e. relationships between item nodes and feature nodes, are referred to as \textit{hyper edges}. 
 It is also worth noting that in the current model we make no assumption on the relationships between
 different feature nodes (i.e. the adjacent matrix on the attribute graph); 
 such connectivities may provide information on the group membership of feature nodes, but not directly on the membership of item nodes, 
 which is the ground truth under concern. 

\section{Bayesian inference on the JSBM}\label{sec:infer}
On the graph $\mathcal{G}$ generated by the JSBM, the problem of \textit{clustering} is defined as recovering ground-true labels $\{t_i^*\}$ exclusively using edge information in the connectivity graph and the attribute graph (i.e., adjacent matrics $\mathbf{A}$ and $\mathbf{F}$); the problem of \textit{semi-supervised classification}, in a slightly different manner, asks to conduct the same recovery but utilizes as extra information a small number of training labels $\{t_{\widetilde i}^*\}$ where $\widetilde i$ belongs to the training set $\Omega$.
    If the parameters $\theta$ in generating the JBSM are known, this classification task can essentially be translated into an inference problem on the group labels $\{t_i^*\}$, which could be viewed as hidden parameters in the model, provided with measurements on a specific type of outcomes (edges in the two graphs) of these parameters $\{t_i^*\}$. Bayesian inference, which amounts to computing the posterior distribution, is typically used for such an inference task. For our problem, under Bayesian rules, the posterior is written as
\begin{equation}\label{eq:bayes}
    P(\{t_i\},\{t_{\mu}\}|\mathcal{G},\theta)=\frac{P(\mathcal{G}|\{t_i\},\{t_{\mu}\},\theta)P_0(\{t_i\},\{t_{\mu}\})}{\sum_{\{t_i\}}P(\mathcal{G}|\{t_i\},\{t_{\mu}\},\theta)P_0(\{t_i\},\{t_{\mu}\}
    )}.
\end{equation}
where $P_0(\{t_i\},\{t_{\mu}\})$ represents prior information on the labels (e.g., information regarding generative parameters $\alpha$ and $\beta$), and $P(\mathcal{G}|\{t_i\},\{t_{\mu}\},\theta)$ is the likelihood of observing graph $\mathcal{G}$ given labels $\{t_i\},\{t_{\mu}\}$ and parameters $\theta$, which is the product-form probability of generating existent edges and hyper edges in $\mathcal{G}$:
\begin{align}\label{eq:likelihood}
    P(\mathcal{G}|\{t_i\},\{t_{\mu}\},\theta)\notag=& \prod_{i}\alpha_{t_i}\prod_{(ij)\in\mathcal{E}}p_{t_i,t_j}\prod_{(ij)\not\in\mathcal{E}}(1-p_{t_i,t_j})\notag\\
    \cdot&\prod_{\mu}\beta_{t_\mu}\prod_{(i\mu)\in\mathcal{F}}q_{t_i,t_{\mu}}\prod_{(i\mu)\not \in\mathcal{F}}(1-q_{t_i,t_{\mu}}).
\end{align}
The clustering problem corresponds to adopting a flat prior, i.e., $P_0(\{t_i\},\{t_{\mu}\})=\textit{constant}$, and semi-supervised classification corresponds to adopting strong prior on item nodes that belong to the training set, such that the probability marginals of (both ground-true and unclassified) item nodes are pinned in the direction of training labels \cite{Zhang2014phase}.


It is well known that computing the normalization of the posterior distribution (i.e., the denominator of (1)) is a \#P problem, thus efficient and accurate approximations are needed for the inference. In the language of statistical physics, the representation of Bayesian inference on the posterior distribution Eq.~$\eqref{eq:bayes}$ corresponds to a Boltzmann distribution at unit temperature: the negative log-likelihood $-\log P(\mathcal{G}|\{t_i\},\{t_{\mu}\}, \theta)$ represents the energy; the normalization constant $\sum_{\{t_i\}}P(\mathcal{G}|\{t_i\},\{t_{\mu}\},\theta)P_0(\{t_i\},\{t_{\mu}\})$ for the posterior is the partition function; the prior information $P_0$ plays the role of external fields acting on item nodes. For the clustering problem with a flat prior, the external field is zero for all nodes; for semi-supervised classification, the external field is infinity for nodes in the training set and zero for unlabelled nodes \cite{Zhang2014phase}.
 
For random sparse graphs, the inference could be studied at the thermodynamic limit using the cavity method from statistical physics~\cite{Mezard2009,yedidia2003understanding}. If the parameters used in generating the SBM is known a priori, the system is on the \textit{Nishimori line}~\cite{nishimori1980exact,iba1999nishimori} and no spin glass phase could appear. Moreover, the replica symmetry cavity method naturally translates into the {\it{belief propagation}} (BP) algorithm, a well-known algorithm for computation on SBM. In BP, cavity messages are passed along directed edges of the factor graph; when the propagation converges, they are used to compute the posterior marginals. Inheriting the message-passing idea, for JSBM, where the factor graph $\mathcal{G}$ is two-fold (i.e., each edge acts as a two-body factor; each hyper edge acts as a multi-body factor), we formulate the iterative equations of BP for three types of messages (see Appendix A for derivations):

\begin{align}
\label{eq:bp}
     &\psi _{t_i}^{i\rightarrow j}=\alpha_{t_i}\frac{e^{-h_{t_i}}}{Z^{i\rightarrow j}}\prod _{k\in \partial i\setminus j}\sum _{t_k}  p_{t_it_k}\psi _{t_k}^{k\rightarrow i}   \prod _{\mu\in\partial   i}\sum_{t_{\mu}}
q_{t_it_{\mu}}\psi_{t_{\mu}}^{\mu\rightarrow i}\notag\\
&\psi _{t_i}^{i\rightarrow \mu}=\alpha_{t_i}\frac{e^{-h_{t_i}}}{Z^{i\rightarrow \mu}}\prod _{k\in \partial i}\sum _{t_k}  p_{t_it_k}\psi _{t_k}^{k\rightarrow i}   \prod _{\nu \in\partial   i\setminus \mu}\sum_{t_{\nu}}
q_{t_it_{\nu}}\psi_{t_{\nu}}^{\nu\rightarrow i}\notag\\
&\psi_{t_{\mu}}^{\mu\rightarrow i}=\beta_{t_{\mu}}\frac{e^{-h_{t_{\mu}}}}{Z^{\mu\rightarrow i}}   \prod _{j\in\partial   \mu\setminus i}\sum_{t_j}
q_{t_{\mu}t_j}\psi_{t_j}^{j\rightarrow \mu}.
\end{align}
Here $\psi_{t_i}^{i\to j}$ are the cavity marginals (messages) passing through item node $i$ to item node $j$, representing the probability of item node $i$ taking label $t_i$ when the item node $j$ is removed from the graph. Similarly, $\psi^{i\to \mu}_{t_i}$ represents the probability of item node $i$ taking label $t_i$ when feature node $\mu$ is removed from the graph, and $\psi_{t_{\mu}}^{\mu\to i}$ represents the probability of feature node $\mu$ taking label $t_{\mu}$ when item node $i$ is removed. $Z^{i\rightarrow \mu}$, $Z^{\mu\rightarrow i}$, and $Z^{i\rightarrow j}$ are normalizing factors; $\partial i$ denotes the set of neighbors of item node $i$ in the graph. In the three equations, variables $h_{t_i}$ and $h_{t_{\mu}}$ are adaptive fields contributed by non-existent edges of the graph, which are formulated as (see Appendix A):
\begin{align}\label{eq:h}
h_{t_i}&=\sum _k\sum _{t_k}p_{t_i t_k}\psi _{t_k}^k+\sum_{\mu} \sum_{t_{\mu}}q_{t_i t_{\mu}}\psi_{t_{\mu}}^{\mu}\notag\\
h_{t_{\mu}}&=\sum_j \sum_{t_j}q_{t_{\mu} t_j}\psi_{t_j}^j
\end{align} 



Once the above iterative equations converge (i.e., messages do not change significantly), using the determined cavity messages, the posterior marginals on the two graphs (connectivity, attribute) could be calculated by:
\begin{align}\label{eq:marginals}
\psi_{t_i}^{i}&=\alpha_{t_i}\frac{e^{-h_{t_i}}}{Z^{i}}\prod _{k\in \partial i} \sum _{t_k}  p_{t_it_k}\psi _{t_k}^{k\rightarrow i}   \prod _{\mu\in\partial i}\sum_{t_{\mu}}q_{t_it_{\mu}}\psi_{t_{\mu}}^{\mu\rightarrow i},\notag\\
\psi _{t_{\mu}}^{\mu}&=\beta_{t_{\mu}}\frac{e^{-h_{t_{\mu}}}}{Z^{\mu}}   \prod _{j\in\partial \mu}\sum_{t_j}
q_{t_{\mu}t_j}\psi_{t_j}^{j\rightarrow \mu}. 
\end{align}
In the end, based on these computed marginals, one is able to estimate the label of each item node, which is the specific label that maximizes the item node's marginal:
\begin{align}\label{eq:argmax}
t_i=\argmax_{t\in\{1,2,...,\kappa\}}\psi_{t_i}^{i}.
\end{align}
In Bayesian inference, $t_i$ is the \textsl{maximum posterior estimate}, representing the optimal result with the minimum mean square error (MMSE)~\cite{iba1999nishimori}. In terms of algorithm design, BP equations essentially adopt the Bethe approximation~\cite{Bethe1935, yedidia2003understanding}, which is a variational distribution:
\begin{equation}
     \label{eq:bethe}
     P(\{t_i\},\{t_{\mu}\}|\mathcal{G},\theta)=\frac{\prod_{ij}\prod_{i\mu}\Phi_{t_i,t_{\mu}}^{i,\mu}\Phi_{t_i,t_{j}}^{i,j}}{\prod_i (\psi_{t_i}^{i})^{|\partial i|-1}\prod_{\mu} (\psi_{t_{\mu}}^{\mu})^{|\partial \mu| -1}},
\end{equation}
where $i,\mu$ represent item nodes and feature nodes. The relationships between variational parameters, the two-point marginals $\Phi_{t_i,t_{\mu}}^{i,\mu}$, $\Phi_{t_i,t_{j}}^{i,j}$(not used in BP equations) and the single-point marginals $\psi_{t_i}^{i}$ (or $\psi_{t_{\mu}}^{\mu}$), are adjusted during the message-passing process in minimizing the Bethe free energy. The core assumption is the conditional independence assumption, which is exact on trees. Thus the Bethe approximation~\eqref{eq:bethe} is always correct for a tree graph in describing joint probabilities, in which case BP algorithms yield exact posterior marginals. Empirically, BP results are shown to be good approximations to true posterior marginals, if the graph is sparse and of locally tree-like structures; hence the BP algorithm is widely applied to inference problems in sparse systems~\cite{Mezard2009}.
 
\section{Detectability transitions of the JSBM}   \label{sec:transition}
For the graph generated by the JSBM with parameters $\theta$, the cavity method provides asymptotically exact analysis, and the belief propagation algorithm (almost) always converges,  due to the Nishimori line property~\cite{nishimori1980exact,iba1999nishimori}. Thus, asymptotically exact properties of the JSBM, such as the phase diagram, can be studied directly at the thermodynamic limit by analysing the messages of the belief propagation.

From (4) and (6), observe that there is a trivial fixed point of BP equations~\eqref{eq:bp} 
\begin{align}\label{eq:para}
\psi _{t_i}^{i\rightarrow j}&=\psi _{t_i}^{i\rightarrow \mu}=\psi_{t_i}^i=\alpha_{t_i}, \notag\\ \psi _{t_{\mu}}^{\mu\rightarrow i}&=\psi _{t_{\mu}}^{\mu}=\beta_{t_{\mu}}.
\end{align}
This fixed point corresponds to the situation where every node in the graph has equal probability of belonging to every group; therefore it is known as the  \textit{paramagnetic fixed point} or \textit{liquid fixed point}. In this case, the marginals do not provide any information about the ground true group labels, whereas only reflect the permutational symmetry of the system.
When this paramagnetic fixed point is stable, the system is in the paramagnetic state, where it is believed that no algorithm can do better than  a random guess in revealing planted group labels. This scenario is known as the non-detectable phase for SBM, whose existence has been mathematically proved in \cite{mossel2018proof}; in this study we extend the analysis to the JSBM with two SBM components. Conceptually, the non-detectable phase in the JSBM is analogous to the ferromagnetic Ising model in paramagnetic phase where the underlying ground-true labels correspond to the all-one configuration, as well as the Hopfield model where underlying ground-true labels refer to the stored patterns. From the viewpoint of statistical inference, edges $\{(ij)\}$ and hyper edges $\{(i\mu)\}$ are observations of the signal (i.e. the ground-true labels), so the paramagnetic phase denotes the situation where the number of observations is too few to reveal any valid information of the signal, such that the system evolves to the paramagnetic fixed point where the label assignment is of equal probability for any node. 

When the number of observations increases, the paramagnetic fixed point will eventually become unstable, leading to a non-trivial fixed point of BP~\eqref{eq:bp} whose values are correlated with the ground true group labels. Where the paramagnetic fixed point of BP becomes unstable indicates the position of the {\textit{detectability transition}} for the JSBM, which posits fundamental limits on the ability of algorithms in revealing information of the ground truth, independent of the specific algorithm being used.

\begin{figure}[h]
\includegraphics[width=0.50\textwidth]{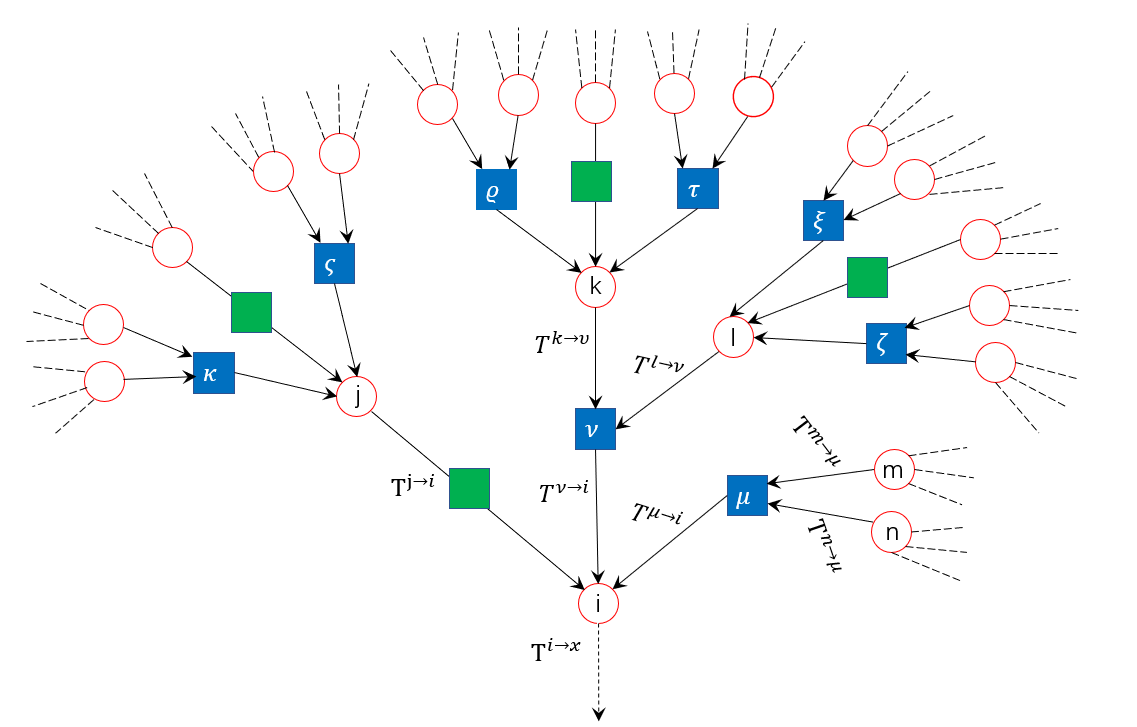}
\caption{Illustration of noise propagation on JSBM. Noises are transmitted from leaves to the root node $i$, with each transmission weighted by eigenvalues of the Jacobian matrices $\mathbf T$, determined by BP message passing equations~\eqref{eq:jacobian}. Red circles denote item nodes, green boxes represent edges connecting item nodes, and blue boxes denote feature nodes.}
\label{fig:noise_pp}
\end{figure}

This phase transition point can be determined by the stability analysis of the paramagnetic fixed point of the BP ~\eqref{eq:para}. Assume that the JSBM graph has $n\to\infty$ item nodes and $m\to\infty$ feature nodes. Each item node is connected to on average $c_1$ item nodes and $c_2$ feature nodes, and each feature node is connected to on average $c_3$ item nodes; when node's degree distribution is Poisson, as in JSBM, $c_1$ and $c_2$ also equal to the average excess degree. Consider putting random noises with zero mean and unit variance on every node (both items and features) of the graph. Assuming a local-tree topology of the graph, after one iteration of the BP equations, the noises will be propagated to on average $c_1$ item nodes through edges, and $c_2c_3$ item nodes through hyper edges (Fig.~\ref{fig:noise_pp}). 
If every leaf node of the tree is associated with random noise of zero mean and unit variance, after $l\to\infty$ iterations of BP equations~\eqref{eq:bp}, the aggregated variance of noises on the root node $i$ can be computed as (see Appendix B for details of derivations)

\begin{equation}
\mathcal{V} = \lim_{l\to\infty}(c_1\lambda_A^2+c_2c_3\lambda_F^4)^l.
\end{equation}
$\lambda_A$ and $\lambda_F$ are the largest eigenvalues of the Jacobian matrices $\mathbf{T}^{i\to j}$ and $\mathbf{T}^{\mu\to i}$, related to the connectivity graph and the attribute graph (i.e. the adjacency matrices $\mathbf{A}$ and $\mathbf{F}$), respectively, which are evaluated at the paramagnetic fixed point (together with a third matrix $\mathbf{T}^{i\to \mu}$):


\begin{align}\label{eq:jacobian}
T^{i\to j}_{t_it_j}&=\frac{\partial \psi _{t_j}^{j\rightarrow x}}{\partial\psi _{t_i}^{i\rightarrow j}} \big|_{\alpha_{t_i}}\,\,\,\,\,\,\,\,=\alpha_{t_i}(\frac{np_{t_i t_j}}{c_1}-1),\notag\\
T^{\mu\to i}_{t_it_{\mu}}&=\frac{\partial \psi _{t_i}^{i\rightarrow x}}{\partial\psi _{t_{\mu}}^{\mu\rightarrow i}} \big|_{\alpha_{t_i},\beta_{t_{\mu}}}=\alpha_{t_i}(\frac{m q_{t_i t_{\mu}}}{c_2}-1),\notag\\
T^{i\to \mu}_{t_{\mu}t_i}&=\frac{\partial \psi _{t_{\mu}}^{\mu\rightarrow j}}{\partial\psi _{t_i}^{i\rightarrow \mu}} \big|_{\alpha_{t_i},\beta_{t_{\mu}}}=\beta_{t_{\mu}}(\frac{m q_{t_{\mu} t_i}}{c_2}-1).
\end{align}

Since $l\to\infty$, the paramagnetic fixed point is unstable under random perturbations whenever $\mathcal{V}>1$. As a result, the detectability phase transition locates at
\begin{align}\label{eq:threshold}
    c_1\lambda_A^2+c_2c_3 \lambda_F^4 =1.
\end{align}
This kind of stability conditions are known in the spin glass literature as  the Almeida-Thouless local stability condition~\cite{de1978stability}), and in computer sciences  as the Kesten-Stigum bound on reconstruction on trees~\cite{kesten1966additional,kesten1967limit}, and the robust reconstruction threshold~\cite{janson2004robust}). 

We demonstrate the phase transition Eq.~\eqref{eq:threshold} through a simple case of JSBM. Consider a $\mathbf P$ matrix with $p_{\textrm{in}}$ being the diagonal and $p_{\textrm{out}}$ being the off-diagonal elements, and a similar $\mathbf Q$ matrix with $q_{\textrm{in}}$ on the diagonal and $q_{\textrm{out}}$ on the off-diagonal. Notice that $p_{\textrm{in}}$ and $p_{\textrm{out}}$, as well as $q_{\textrm{in}}$ and $q_{\textrm{out}}$, are related by:
\begin{align}
p_{\textrm{in}}+(\kappa-1)p_{\textrm{out}}&=\frac{\kappa c_1}{n},\nonumber\\ q_{\textrm{in}}+(\kappa-1)q_{\textrm{out}}&=\frac{\kappa c_2}{m},
\end{align}
hence there is only one free parameter for each matrix. We introduce $\epsilon_1=\frac{p_{\textrm{in}}}{p_{\textrm{out}}}$ and $\epsilon_2=\frac{q_{\textrm{in}}}{q_{\textrm{out}}}$. For this simple JSBM, the first eigenvalues of the two Jacobian matrices are expressed as:
\begin{align}
    \lambda_A = \frac{1-\epsilon_1}{1+(\kappa-1)\epsilon_1} &&
    \lambda_F = \frac{1-\epsilon_2}{1+(\kappa-1)\epsilon_2}.
\end{align}

Numerical experiments are conducted to verify our theoretical results. The performance of BP is evaluated using the overlap $\mathcal{O}$ between the BP results $\{t_i\}$ obtained from Eq.~\eqref{eq:argmax} and the ground truth $\{t_i^*\}$:
\begin{equation}\label{eq:ovl}
\mathcal{O}(\{t_i\},\{t_i^*\}) = \max_{\pi}
\frac{1}{n} \sum_{i=1}^{n}\delta_{\pi(t_i),t_i^*},
\end{equation}
which is maximizing over all permutations of groups $\pi$ (i.e., $|\pi| = \kappa !$). In Eq.~\eqref{eq:ovl}, if the inferred labels $\{t_i\}$ are chosen randomly which has nothing to do with the ground truth $\{t_i^*\}$, the overlap $\mathcal{O} = 1/\kappa$ (lower bound); if there is an exact match between $\{t_i\}$ and $\{t_i^*\}$, $\mathcal{O} = 1$ (upper bound). For a small number of groups, overlap is a commonly used metric for estimating the similarity between two group assignments. When the number of groups is large, or with different group sizes, more advanced measures are expected, such as the normalized mutual information (NMI) and its variances~\cite{danon2005comparing,rnmi}. 

\begin{figure*}[tb]
\centering
\subfigure[]{
\includegraphics[width=0.31\textwidth]{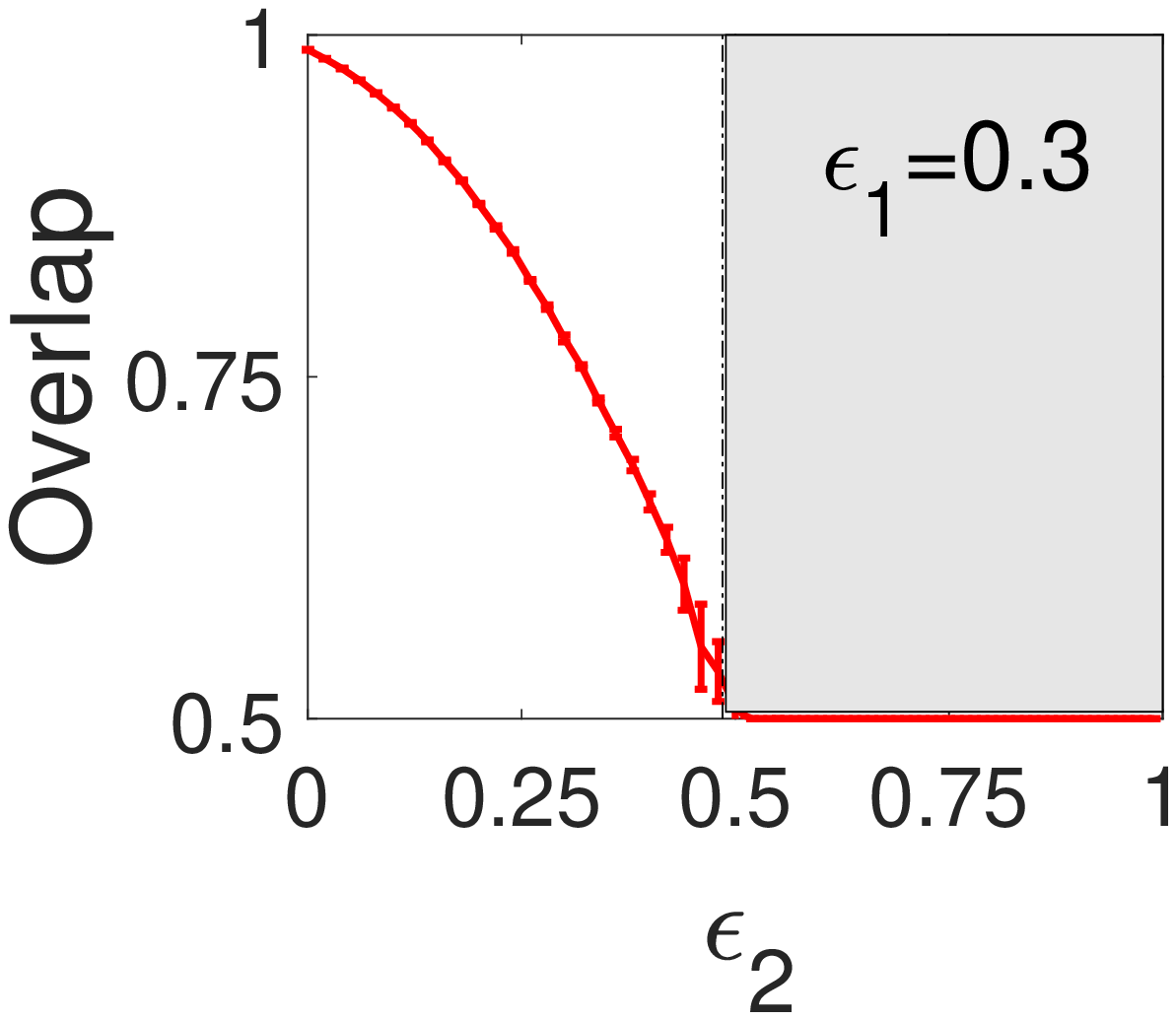}
}
\subfigure[]{
\includegraphics[width=0.31\textwidth]{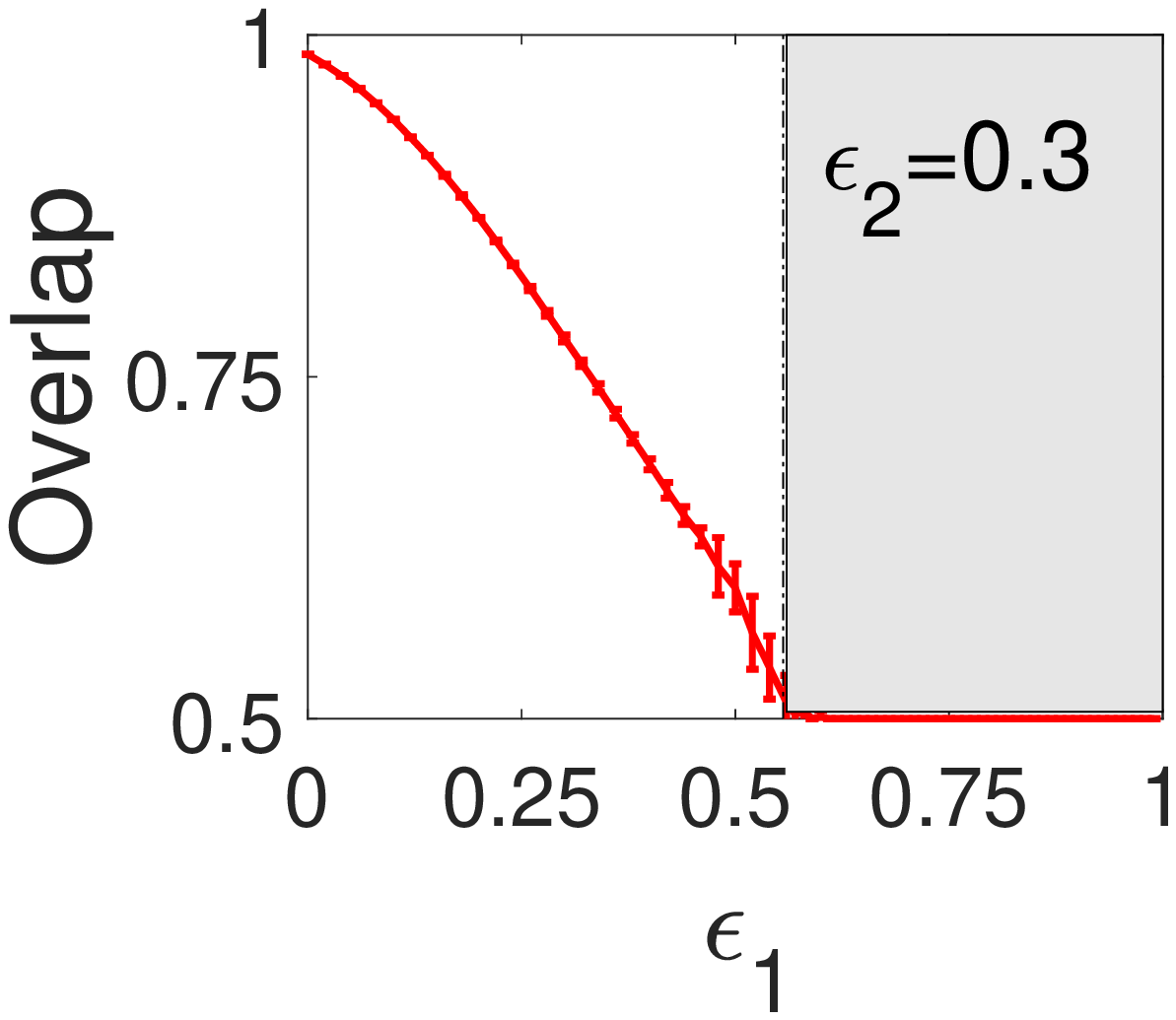}
}
\subfigure[]{
\includegraphics[width=0.31\textwidth]{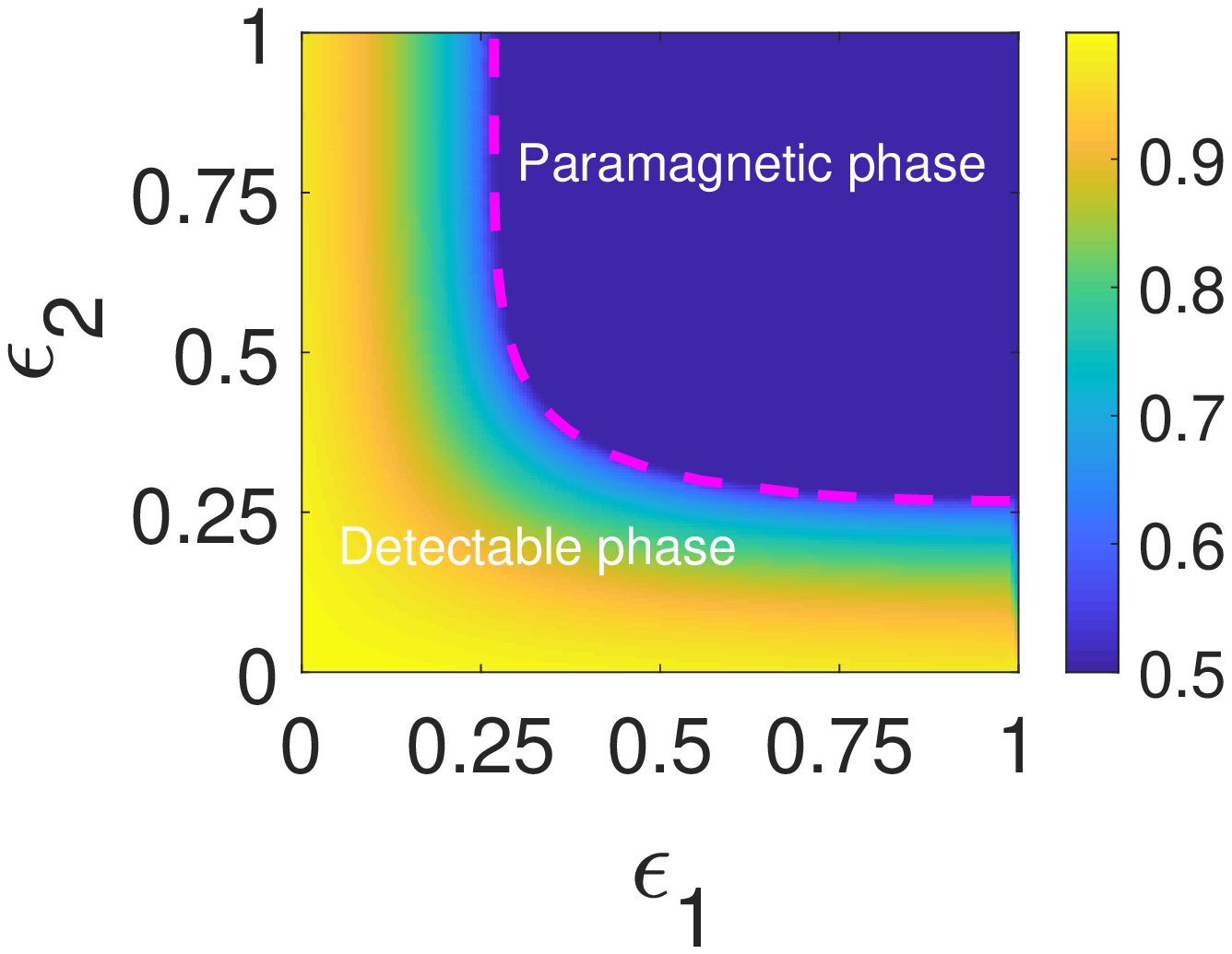}
}
\caption{Detectability phase transition of JSBM. Synthetic JSBM graphs are generated with $n=2*10^5$,$m=2*10^5$, group numbers $\kappa=2$, and average degree $c_1=c_2=3$. (a) $\&$ (b) Overlap as a function of $\epsilon_1$ or $\epsilon_2$, with the other parameter fixed to 0.3 ((a) $\epsilon_1$ fixed; (b) $\epsilon_2$ fixed). The numerical detectability phase transition point agrees with the theoretical calculation \eqref{eq:threshold}, except for small fluctuations due to the finite size effect. (c) Overlap on the $\epsilon_1$ -- $\epsilon_2$ plane. Dashed line indicates the theoretical detectability phase transition point, separating the detectable phase from the paramagnetic phase. Numerical and analytical results are consistent.  \label{fig:ovl}} 
\end{figure*}

Results are shown in  Fig.~\ref{fig:ovl}. In Figure 3(a), $\epsilon_1$ is fixed at 0.3 and $\epsilon_2$ varies; in Figure 3(b) it is the other way around. In both Figure 3(a) and 3(b), the overlap between BP results and the (synthetic) ground truth are plotted against the varied $\epsilon$ ($\epsilon_1$ or $\epsilon_2$), which are optimal in the thermodynamic limit among using all possible algorithms. The overlap is close to $1.0$ for small $\epsilon$, indicating near-perfect reconstructions of the ground truth; with increased values of $\epsilon$, the accuracy of BP inference decreases, which eventually downgrades to $0.5$. This is consistent with the theoretical limit of the detectability phase transition Eq.~\eqref{eq:threshold} (indicated by dashed lines). For a large $\epsilon$, the system goes beyond the phase transition point and lies in the paramagnetic phase, where the overlap is always $0.5$, i.e., results of the BP detection are indistinguishable from that of a 50-50 random guess. In Figure 3(c), the accuracy of BP for JSBM is shown on the $\epsilon_1$-$\epsilon_2$ plane, where the overlap decays from large values (yellow) to the non-informative $0.5$ (blue). The dashed line represents the theoretical phase transition point, consistent with the numerical results.

\section{from belief propagation to graph convolution network}\label{sec:bpgcn}
When applying BP algorithms (Eq.(4)) to real-world graphs or synthetic graphs without a priori knowledge of the generative process, a critical problem is to determine the parameters $\theta$ of the underlying JSBM. For a pure clustering problem, the classical approach for this task is using the Expectation Maximization (EM) algorithm ~\cite{dempster1977maximum}, in which parameters are updated through maximizing the total log-likelihood of data (i.e., minimizing the total free energy). In practice, however, the EM approach is prone to overfitting~\cite{decelle2011asymptotic} and may often be trapped in local minima. A different approach could be taken if a small number of ground truth labels are available, as is the setting of the current study, since now the parameters could be determined in a semi-supervised fashion. Such a semi-supervised update of parameters could possibly be achieved through back-propagation on a neural network structure, which also facilitates the message passing of BP since it needs to be done in an iterative manner. With this idea in mind, we propose a solution framework for the JSBM which contains two steps in each epoch: in the forward step, BP equations propagate to finite time steps (layers) and conduct messages passing, and in the backward step, the parameters $\theta$ of the underlying JSBM are determined in a supervised approach, and back-propagate to BP layers. 

Essentially, the graph convolution network (GCN) ~\cite{kipf2017semi} structure is adopted in the above framework, which is an outstanding neural network model that triggers a large number of variants (see below). On a GCN, values on graph nodes $\mathbf X$ propagate forward among layers; in a high-level description, the propagation takes the following form: 
\begin{equation}\label{eq:gcn1}
\mathbf X_{l+1}=\sigma(\mathbf U \mathbf X_{l}\mathbf W),
\end{equation}

where $\sigma(\cdot)$ represents an activation function, $\mathbf X_{l+1}\in \mathbb{R}^{n\times \kappa_{l}}$ is the neural network state at layer $l$ ($\kappa_{l}$ denoting the dimension of state at layer $l$, which is not necessarily equal to $\kappa$ in the GCN). The matrix $\mathbf W\in\mathbb{R}^{\kappa_{l+1},\kappa_l}$ is the trainable weight matrix at the $l$-th layer, and the propagator $\mathbf U$ is responsible for convolving the neighborhood of nodes, a kernel shared over the entire graph. 

Based on the GCN structure, we propose the \textit{Belief Propagation Graph Convolution Network (BPGCN)}, as our solution framework of the JSBM. Marginals and cavity messages $\psi$ are network states; kernels take the job of the products and summations in Eq.~\eqref{eq:bp} and \eqref{eq:marginals}; and parameters of the JSBM $\mathbf P$ and $\mathbf Q$ become the weight matrices of the neural network, which are to be gradually learnt through back-propagation. Under these conventions, the propagation from the $l$-th layer to the ${l+1}$-th layer on a BPGCN is formulated as:
\begin{align}
   \label{eq:pbp}
&\mathbf \Psi_{l+1}^{\mu}=\mathrm{Softmax}\left[ \mathbf{U_I}\log(\mathbf\Psi^{i\to \mu}_{l}\mathbf P)\right]\notag\\
      &\mathbf\Psi^i_{l+1}=\mathrm{Softmax}\left[ \mathbf U_{II}\log(\mathbf\Psi^{\mu \rightarrow i }_l\mathbf Q)+\mathbf U_{III}\log(\mathbf\Psi^{i \rightarrow j}_l\mathbf P)\right]\notag\\
    &\mathbf \Psi^{i\rightarrow j}_{l+1}=\mathrm{Softmax}\left[\mathbf B_I\text{log}(\mathbf\Psi^{\mu \rightarrow i }_l\mathbf Q)+\mathbf B_{II}\log(\mathbf \Psi^{i \rightarrow j}_l\mathbf P)\right]\notag\\
    &\mathbf\Psi^{i\rightarrow \mu}_{l+1}=\mathrm{Softmax}\left[ \mathbf B_{III}\log(\mathbf \Psi_l^{i \rightarrow j}\mathbf P)+\mathbf B_{IV}\log(\mathbf \Psi^{\mu\rightarrow i}_l \mathbf Q)\right]\notag\\
    &\mathbf\Psi^{\mu\rightarrow i}_{l+1}=\mathrm{Softmax}\left[\mathbf B_{V}\log(\mathbf\Psi^{i \rightarrow \mu }\mathbf Q)\right]
\end{align}
where $\mathrm{Softmax}( z_t)= \frac{e^{ z_t}}{\sum_{s=1}^\kappa e^{z_s}}$ is used as the activation function, inherited naturally from BP which asks to normalize marginal probabilities with $\kappa$ components. If $\kappa=2$, the $\mathrm{Softmax}$ activation function reduces to the Sigmoid function. Also note the logarithm on $\psi$ inside the Softmax, which might be considered as a part of the activation function in a strict sense. Kernel matrices $\mathbf{U_I}$ to $\mathbf{U_{III}}$, $\mathbf{B_I}$ to $\mathbf{B_V}$ are non-backtracking matrices~\cite{krzakala2013spectral} that encode adjacency information of cavity messages and marginals (see Appendix D for details). In practice, random values are used in the input layer of the network, for marginals $\mathbf \Psi^i_0\in \mathbb R^{n\times \kappa}$ and $\mathbf \Psi^{\mu}_0\in \mathbb R^{m \times \kappa}$, as well as messages $\mathbf \Psi^{i\rightarrow j}_0 \in \mathbb R^{2M_A\times \kappa}$, $\mathbf \Psi^{i\rightarrow \mu }_0\in \mathbb R^{2M_F\times \kappa}$ and $\mathbf \Psi^{\mu\rightarrow i }_0\in \mathbb R^{2M_F\times \kappa}$, where $M_A$ and $M_F$ are the number of edges (in the connectivity graph) and hyper-edges (in the attribute graph), respectively. 

The marginals in the last layer of BPGCN $\mathbf\Psi=\{ \mathbf\Psi^{i}_L  \}\in\mathbb [0,1]^{n\times \kappa}$ are the output of a $L-$layer BPGCN. A loss function on the fraction of marginals belong to training labels is adopted for the supervised learning. A common choice of the loss function for classification is the cross entropy $\mathcal{L}$; in our model, it is defined as:
\begin{equation}
\label{eq:loss}
\mathcal{L} = -\sum_{i\in \Omega}\sum_{s=1}^\kappa (\mathbf y_{i})_s\ln (\Psi^{i}_L)_s
\end{equation}
where $\Omega$ denotes the training set of item nodes (a fraction of ground-true nodes), $\mathbf y_{i}=\{0,0,..,1_{position(t_i^*)},...0\}$ is a one-hot vector denoting the ground-truth label of node $i$ in the training set.

Training the BPGCN is the same as training other GCNs, as mentioned earlier: in each epoch, we first do a forward pass to obtain the computed marginals in the last layer, based on which we calculate the loss function on the training set; after that, we use the \textit{Back Propagation}~\cite{goodfellow2016deep} algorithm to compute the gradients of the loss function with respect to elements in $\mathbf P$ and $\mathbf Q$, and then apply (stochastic) gradient descent or its variants (e.g. ADAM \cite{kingma2014adam}) to update the parameters. The iteration stops when the results converge, or after a finite number of epochs. In the end, the performance of the BPGCN algorithm is evaluated by the accuracy (i.e., overlap, Eq.(14)) of label assignments on the ground-true item nodes in the test set.

A critical difference between BPGCN and traditional BP (including the semi-supervised version~\cite{Zhang2014phase}) is that, BP minimizes the (Bethe) free energy, whereas BPGCN minimizes the loss function evaluated on the training data. On JSBM synthetic graphs with matched parameters, the free energy is theoretically the best loss function to minimize; however, on graphs not generated by JSBM, minimizing the (free) energy is prone to overfitting~\cite{decelle2011asymptotic,Zhang2014pnas}. More importantly, minimizing the loss function is more flexible since the function could be formulated in different ways, for example, new (informative, or regularization) terms could be added. Notably, one implicit feature of inference on JSBM is that, the classification on feature nodes (i.e., $\Psi^{\mu}_L$) is left unconstrained, in both BP (equation (6)) and BPGCN (equation (17)), since the ground truth of feature labels are often not available. Nevertheless, in specific situations, such constraints could be readily incorporated in the loss function. For example, sometimes it is reasonable to believe that, the distribution of classified feature labels should be close to a uniform distribution (or Gaussian in other cases); thus in these cases, we are able to append a term in the loss function to constrain the classification performance on features. In particular, using a one-hot vector $\mathbf w_{\mu}=\{1,1,..,1\}$ to characterize the uniform distribution, we could formulate a new loss function as:
\begin{equation}
\label{eq:loss}
\mathcal{L}' = -\sum_{i\in \Omega}\sum_{s=1}^\kappa (\mathbf y_{i})_s\ln (\Psi^{i}_L)_s - \eta\sum_{\mu\in \Omega_m}\sum_{s=1}^\kappa (\mathbf w_{\mu})_s\ln (\Psi^{\mu}_L)_s
\end{equation}
where $\eta$ is a damping factor, and $\Omega_m$ denotes the set of all feature nodes successfully classified. This new loss function may probably help enhance the classification performance of BPGCN on real-world networks, although it is certainly not optimal for synthetic JSBM graphs with nonzero $\eta$. It is also worth noting that the adaptive fields (Eq.~\eqref{eq:h}), which in traditional BP are contributed by non-edges and are indispensable, are not necessary in BPGCN, because the traning labels automatically balance the group sizes. 

Comparing BPGCN with canonical GCNs, two major differences emerge. First, the activation function in BPGCN (Softmax) is not chosen arbitrarily, but rather determined by BP message passing equations; this is in contrast with common GCNs where the activation function may take various forms, such as \text{ReLU}, \text{PReLU} or \text{Tanh}, but without sufficient physical or mathematical warrants. Second, there are only few parameters to be trained in BPGCN, which are the elements of $\mathbf P$ and $\mathbf Q$, and they are shared across all layers of the neural network. Although this may narrow the overall representational power of BPGCN, the problem of overfitting could nevertheless be obviated to a great extent. Indeed, in semi-supervised classification, the amount of training data is often much less than that of  (completely) supervised classification, whereas the number of observations, in the case of networks, may be proportional to the number of edges (and hyper edges in our model); therefore, the sharing of parameters across layers is believed to be extremely helpful in preventing overfit of the training data.


\section {Comparing BPGCN with other graph convolution networks} \label{sec:comparing} 

In this section we compare the classification performance of BP and BPGCN, constructed on the joint stochastic block model, with several state-of-the-art graph convolution networks, including the \textit{(standard) GCN}, the \textit{Approximate Personalized Propagation of Neural Predictions (APPNP)}, the \textit{Graph attention network (GAT)}, and the \textit{Simplified Graph Convolution Network (SGCN)}.

\noindent $\mathbf{1}$. {\it{(standard) Graph Convolution Network (GCN)}} ~\cite{kipf2017semi} \\
The standard GCN is probably the most famous graph convolution network, which drastically outperformed all non-neural-network algorithms when first proposed in $2017$.
The forward propagation rule of standard GCN is formulated as
\begin{equation}
  \label{eq:gcn}
    \mathbf H^{(l+1)}=\mathrm {ReLU}\left(\widetilde{\mathbf A}\mathbf H^{(l)}\mathbf W^{(l)}\right),
\end{equation}
where $\mathbf H^{(l)}$ and $\mathbf W^{(l)}$ are states of hidden variables and weight matrices at the $l$-th layer. $\widetilde{\mathbf{A}}$ is the graph convolution kernel, defined as
\begin{equation}\label{eq:A}
\widetilde{\mathbf A}=\mathbf D^{-\frac{1}{2}}(\mathbf A+\mathbf I)\mathbf D^{-\frac{1}{2}}.
\end{equation}
$\mathbf D$ is a diagonal degree matrix with $D_{ii}=1+\sum_k A_{ik}$.
The propagation rule of the standard GCN was motivated by a first-order approximation of localized spectral filters (see Appendix C). Hence a two-layer standard GCN is written as:
\begin{equation}
\label{eq:twogcn}
 \mathbf{ Z}=\mathrm{Softmax}[\widetilde{\mathbf A}\text{Relu}(\tilde{\mathbf A}\mathbf R\mathbf W^0)\mathbf W^1],
\end{equation}
where $\mathbf R$ is the encoded information and $\mathbf Z\in\mathbb R^{n\times \kappa}$ is the classification result.

\noindent $\mathbf{2}$. {\it{Approximate Personalized Propagation of Neural Predictions (APPNP)}}~\cite{klicpera2018combining}\\
APPNP extracts feature information (encoded in $\mathbf R$) to hidden neuron states $\mathbf H$ using a multi-layer perceptron (MLP):
\begin{equation}
\mathbf H=\mathbf Z^{(1)}=\mathrm{MLP}(\mathbf R).
\end{equation}
Similar to the GCN, the hidden states $\mathbf H$ are propagated via the personalized PageRank scheme to produce predictions of node labels:
\begin{align}
\label{eq:appnp}
&\mathbf Z^{(k+1)}=(1-\omega) \widetilde{\mathbf A} \mathbf Z^{(k)}+\omega \mathbf H \notag\\
&\mathbf{Z}=\mathrm{Softmax}[(1-\omega) \widetilde{\mathbf A} \mathbf Z^{(K-1)}+\omega \mathbf H],
\end{align}
where $\mathbf Z$ are explicit components of the hidden state $\mathbf H$ of each layer, and $\omega \in [0,1]$ is a weight factor. In the recent benchmarking study on the performance of various GCNs, APPNP produces the best results on multiple datasets~\cite{Fey/Lenssen/2019}.

\noindent $\mathbf{3}$. {\it Graph attention network (GAT)} \cite{2018graph} \\
GAT adopts the \textit{attention mechanism}~\cite{vaswani2017attention} in which attention coefficients between pairs of connected nodes are regarded as weights. This scheme can be viewed as based on the adjacency matrix of the graph, but with adjustable weights on edges.

\noindent $\mathbf{4}$. {\it Simplified Graph Convolution Networks (SGCN)~\cite{pmlr-v97-wu19e} }
SGCN tries to remove redundant and unnecessary computations from GCN by adopting a low-pass filter followed by a linear classifier. As a result,  SGCN takes a simplified propagation rule using the $k$-th power of the adjacency matrix $\widetilde {\mathbf A}$, as in GCN:
\begin{align}
  \label{eq:sgcn}
   \mathbf Z=\mathrm{Softmax}(\widetilde{\mathbf A}^k \mathbf R\mathbf W),
\end{align}
where $\mathbf R$, $\mathbf W$, $\mathbf Z$ are feature information, weights and classification results, respectively. Empirical results show that the simplification process may yield positive impacts on the accuracy of GCN, and could dramatically speed up the computation.

\subsection{Results on synthetic networks}

First we compare these GCN algorithms on synthetic networks, generated by the JSBM. Each JSBM graph consists of $n=10,000$ item nodes, $m=10,000$ features nodes, and $\kappa = 5$ groups; each item node has on average $c_1$ neighbors in the connectivity graph and $c_2$ neighbors in the attribute graph; each feature node is connected to on average $c_3$ item nodes. In the semi-supervised experiments, 5$\%$ randomly chosen item nodes with ground-true labels are used as the training set, and another 5$\%$ nodes are used as the validation set. The rest of nodes belong to the test set, on which the accuracy of the classification results is evaluated, by the measure of overlap (Eq. (14)). For BP and BPGCN, diagonal matrices are used as the initial values for $\mathbf P$ and $\mathbf Q$, with $p_{\textrm{in}}$ (and $q_{\textrm{in}}$) on the diagonal and $p_{\textrm{out}}$ (and $q_{\textrm{out}}$) on the off-diagonal entries. Define ratio $\epsilon_1={p_{\textrm{out}}}/{p_{\textrm{in}}}$ and $\epsilon_2={q_{\textrm{out}}}/{q_{\textrm{in}}}$, both of which could be viewed as the (inverse) signal-to-noise ratio; for all synthetic networks, we use $\epsilon_1=\epsilon_2=0.5$ as initial values  for BPGCN. While, for BP we use the right initial values which are used to generate corresponding graphs, naturally obtaining asymptotically optimal results. The initial values of $\epsilon_1$ and $\epsilon_2$ for BPGCN could be instead determined in a soft manner by the validation set; however, results show that this treatment is not very  necessary when compared to BP.

\begin{figure*}[tb]
\centering
\subfigure[]{
\includegraphics[width=0.31\textwidth]{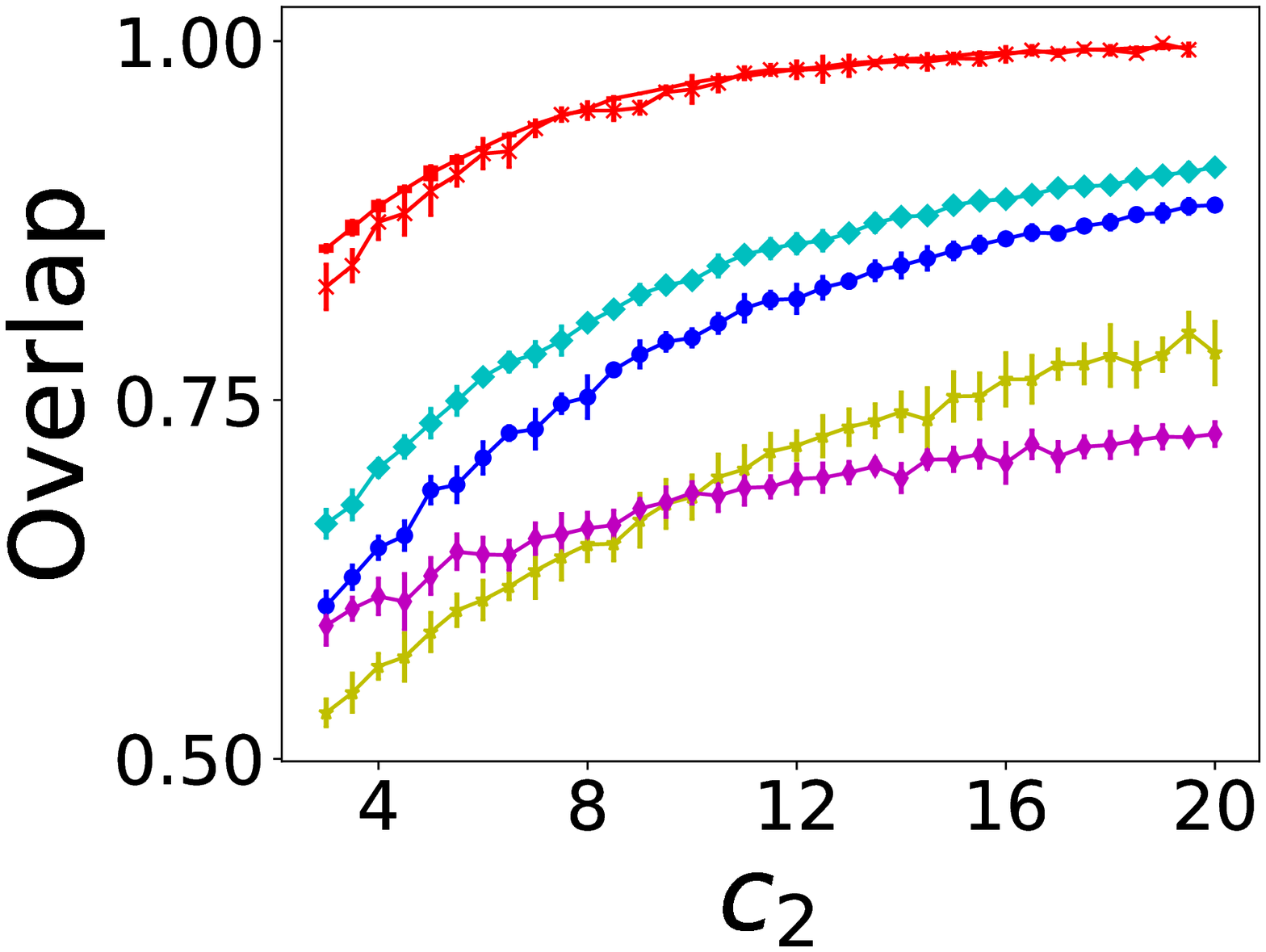}
}
\subfigure[]{
\includegraphics[width=0.31\textwidth]{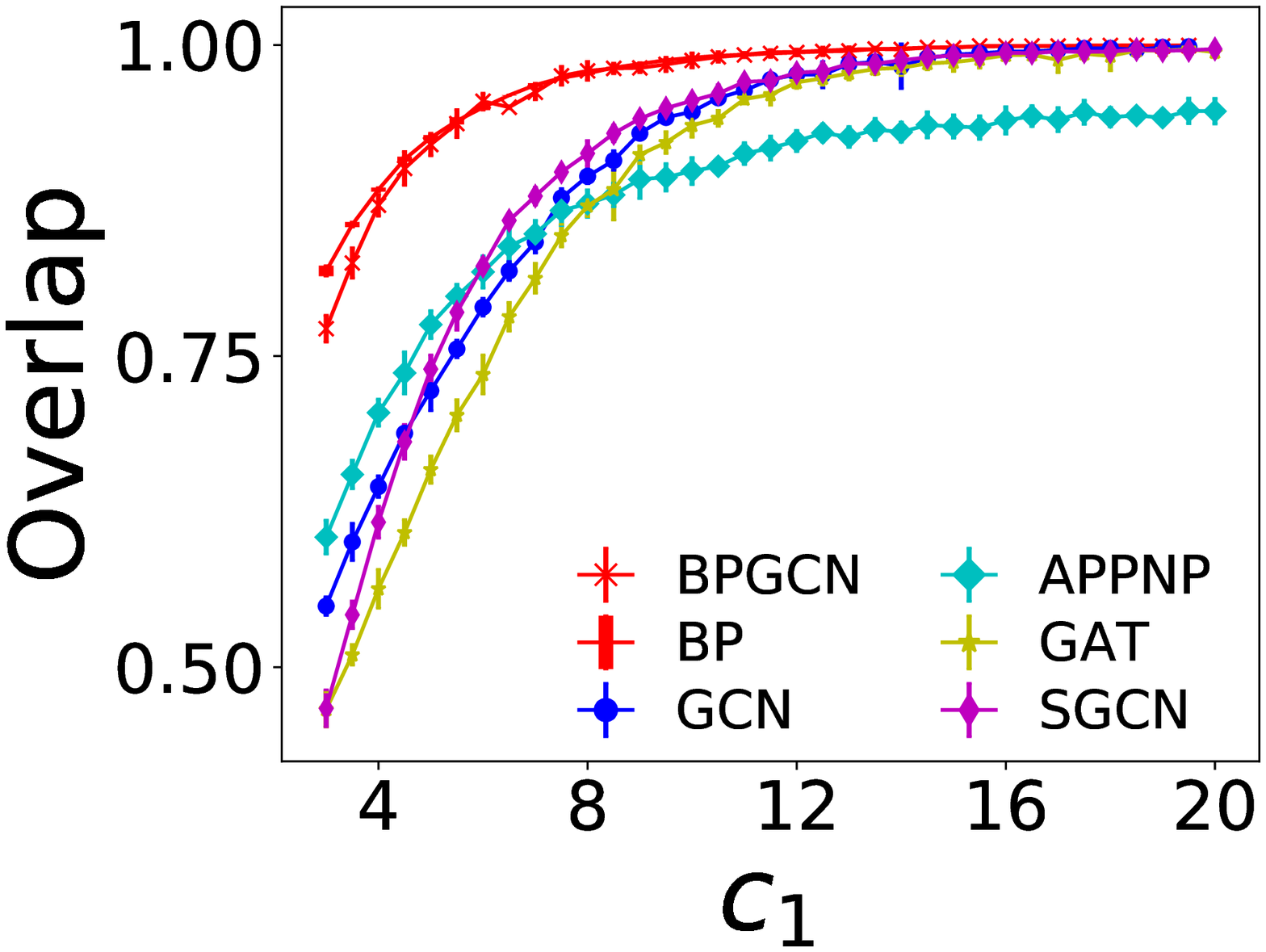}
}
\subfigure[]{
\includegraphics[width=0.31\textwidth]{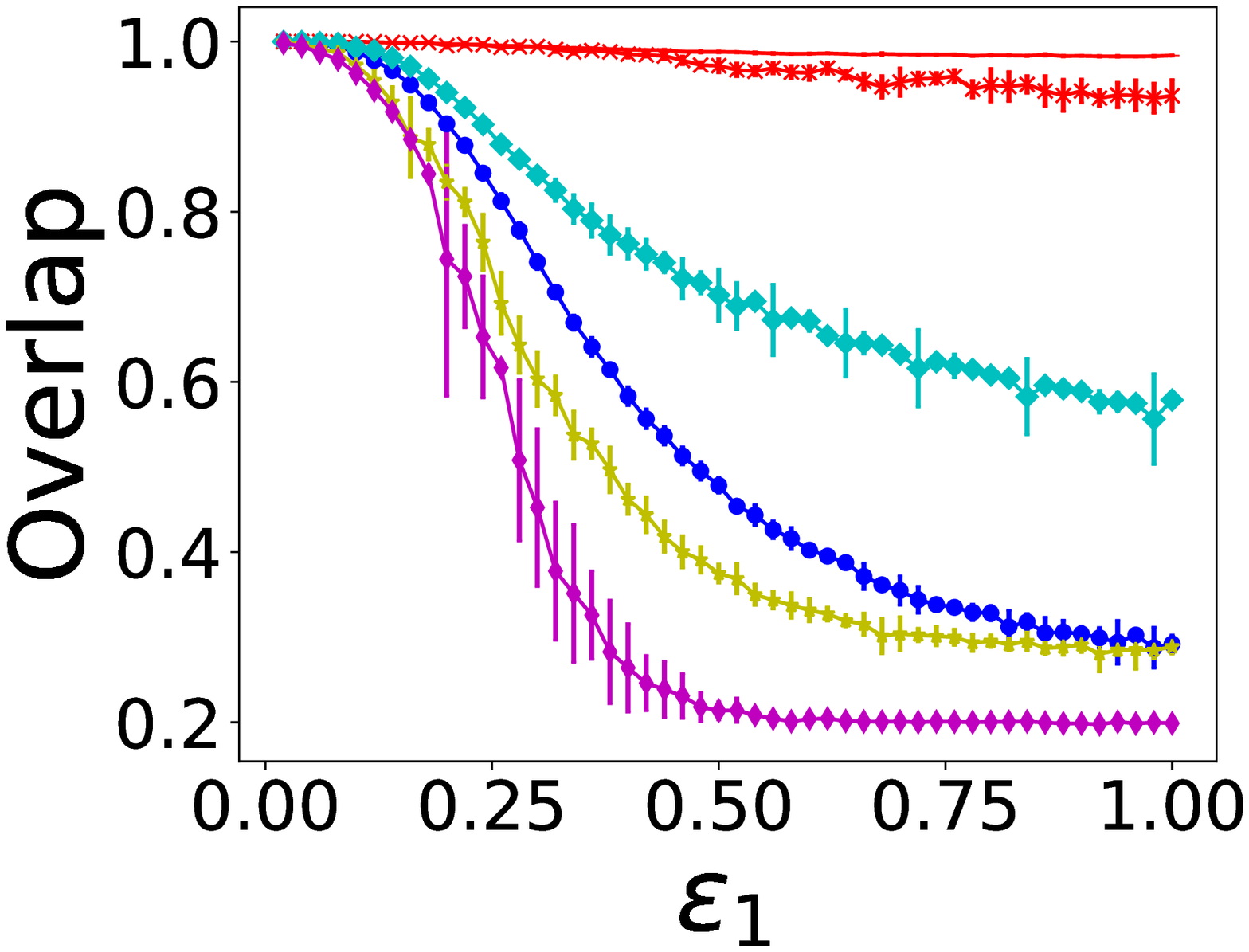}
}
\caption{Comparing BP and BPGCN with other GCNs on synthetic networks. Networks are generated by JSBM with $n=10,000$ item nodes, $m=10,000$ features, and $\kappa=5$ groups. During classification, $5\%$ randomly selected nodes are used as the training set, another $5\%$ used as the validation set, and the rest of nodes are the test set (all synthetic nodes have ground-true group labels), on which the metric of overlap ~\eqref{eq:ovl} is evaluated to indicate different algorithms' performance. Each data point in the figure is averaged over $10$ instances. (a) $\epsilon_1=0.1$, $\epsilon_2=0.2$, $c_1=4$, and $c_2$ varies; (b) $\epsilon_1=0.1$, $\epsilon_2=0.2$, $c_2=4$, with $c_1$ varies; (c) $c_1=c_2=10$, $\epsilon_2$ is fixed to 0.2, and $\epsilon_1$ ranges from $0$ to $1$. \label{fig:ovljsbm}} 
\end{figure*}

Extensive numerical experiments have been carried out on large synthetic graphs with varying average degrees and varying $\epsilon_{1,2}$ (Fig.~\ref{fig:ovljsbm}; see also Appendix E). First, it is confirmed that BPGCN results are perfectly aligned with the (asymptotically) optimal BP results , outstanding all other GCNs. In Figure 4(a), $c_1$ is fixed to $4$, and $c_2$ increases from $3$ to $20$. We can see that all other algorithms (GCN, APPNP, GAT, and SGCN) work much worse than BPGCN, even when $c_2$ is quite large. In Figure 4(b), $c_2$ is fixed to $4$ and $c_1$ is varied. Similar to Figure 4(a), it shows that when $c_1$ is small, GCN, APPNP, GAT, and SGCN perform far worse than BP and BPGCN. These results on synthetic graphs clearly show the superiority of BPGCN over existing popular GCN algorithms; the reason to account for the poor performance of the reference GCN algorithms is due to the \textit{sparsity} of the graph (see below).

In Figure 4(c), $\epsilon_2$ is fixed to $0.1$; the average degrees of the synthetic graph are fixed as $c_1=10$ and $c_2=10$; $\epsilon_1$ is varied. Quite surprisingly, results show that BPGCN works perfectly in the whole range of $\epsilon_1$, even when $\epsilon_1$ is close to 1, i.e., the $\mathbf P$ matrix is homogeneous, whereas conventional GCNs quickly fail when $\epsilon_1$ increases. This phenomenon indicates that conventional GCNs under discussion have great difficulties in extracting the information on group labels from features, when the group structures of the graph are noisy, as one would imagine; nevertheless, to an extraordinary extent, BPGCN is not subject to this failure. Further check on the outputs of conventional GCNs demonstrate that these results all have good overlap in the training set, but poor overlap in the test set, indicating clear \textit{overfitting} to  training labels. In what follows, we discuss the two issues (the \textit{sparsity} issue and the \textit{overfitting issue}) emerged in the results and analyze how they are successfully overcome by BPGCN.

\textbf{\textit{The sparsity issue---}}
Properties of the forward-propagation of GCNs are closely related to their linear convolution kernels, hence the reason to account for the sparsity issue can be uncovered by studying the spectrum of the linear kernels used in graph convolutions. In GCN, SGCN and APPNP, the linear convolution kernel is a variant of the normalized adjacency matrix $\widetilde{\mathbf A}$~\eqref{eq:A}. It has been established in e.g.~\cite{krzakala2013spectral,zhang2016robust} that this type of linear operators have localization problems on large sparse graphs, with leading eigenvectors only encoding local rather than global information on group structures (Appendix C). In contrast, our BPGCN is immune to the sparsity issue, because the convolution kernels of BPGCN are the non-backtracking matrices, which naturally overcome the localization problem on large sparse graphs~\cite{krzakala2013spectral}. Therefore, inspired by BPGCN, a straightforward way of overcoming the sparsity issue in classic GCNs might be to consider using a linear kernel that does not trigger the localization problem in sparse graphs, such as the non-backtracking matrix or the X-Laplacian~\cite{zhang2016robust}. 

\textbf{\textit{The overfitting issue---}}
From Eq.~\eqref{eq:gcn}, ~\eqref{eq:appnp} and ~\eqref{eq:sgcn}, one could observe that in these models the linear filters always operate directly on the weight matrices or on the hidden states. This implies an underlying assumption of conventional GCNs: the relational data (i.e. edges, or the adjacency matrix of the connectivity graph) must always contain information on item nodes' group structures (i.e., information on nodes' labels). This is a natural assumption, yet may not always be true. In an advanced manner, BPGCN relaxes this assumption by learning an affinity matrix $\mathbf P$ that essentially stores the learned signal-to-noise ratio ($\epsilon_{1,2}$) indicating the distribution of edges; hence it may identify that there is not necessarily information encoded in edge connectivities.

\subsection{Results on real-world datasets}
Next we apply BPGCN and reference algorithms on several well-known real-world networks with and without node features. The \net{Karate Club} network~\cite{zachary1977information} and \net{Political Blogs} network \cite{adamic2005political} are classical network datasets containing community structures. Since in these two networks there is no feature of nodes, canonical GCNs commonly use the identity matrix as the feature matrix $\mathbf{F}$~\cite{kipf2017semi}. Given their small sizes, for the \net{Karate Club} network we use $2$ nodes per group as training nodes and $2$ nodes per group as validation nodes, and for the \net{Political Blogs} network we use $10$ nodes per group for training and $10$ nodes per group for validation. 
The \net{Citeseer}, \net{Cora} and \net{Pubmed} networks are standard datasets for semi-supervised classification widely used in graph neural network studies; we follow the splitting rule of training, test and validation sets introduced in \cite{kipf2017semi} on these graphs. For all these networks, there are ground-true labels for nodes' group membership, coming from either public information (for \net{Karate club} and \net{Political blogs} networks) or empirical expert knowledge (i.e., research areas of articles in citation networks). Same as the case for synthetic data, the performance of tested algorithms is evaluated by the overlap between classification results and the ground-true labels in the test set.

For BPGCN, a tunable external field is adopted and corresponding hyper parameters are introduced, so as to adjust the relative strength of the training label on nodes. Identical to the case in synthetic graphs, diagonal matrices are used as initial values for $\mathbf P$ and $\mathbf Q$, controlled by the two signal-to-noise ratios $\epsilon_1={p_{\textrm{out}}}/{p_{\textrm{in}}}$ and $\epsilon_2={q_{\textrm{out}}}/{q_{\textrm{in}}}$. Unlike on synthetic graphs where initial $\epsilon_1$ and $\epsilon_2$ are fixed at 0.5, on real-world graphs we did a coarse search using the validation set to determine  proper values for $\epsilon_1$ and $\epsilon_2$ during the preprocessing step, together with the search for proper hyper parameters, including the strength of the external field strength and the number of layers of BPGCN (see Appendix D). 

Classification results are shown in the Table~\ref{tab:results}. On the two networks without node features, all tested GCN algorithms perform quite well, and label information is successfully extracted from edge connectivities. BPGCN outperforms other GCNs for the \net{Polical blogs} network; the good performance may come from its non-backtracking convolution kernel, which has good spectral properties on large sparse graphs such as the \net{Political blogs} network~\cite{krzakala2013spectral}. On networks with node features, the comparison of classification results is less straightforward. On \net{Citeseer} and \net{Cora}, the performance of BPGCN is certainly comparable to other GCNs, and in both cases it is superior to the performance of at least one reference algorithm. Nevertheless, BPGCN works poorly on the \net{Pubmed} network, yielding the worse performance among all tested GCNs. The reason is that the \net{Pubmed} network contains $19717$ item nodes, but only $500$ features; when features are densely connected to item nodes, the attribute graph significantly deviates from being a sparse random graph, a critical assumption our BPGCN algorithm and the underlying JSBM rely on. Indeed, it is verified that, when completely ignoring features and only using edge connectivities in classification, BPGCN yields a classification accuracy at $71.0\%$, better than $70.0\%$. Since the Multilayer percetron (MLP) method which only uses the feature information already achieves a high accuracy, a possible remedy for BPGCN in situations where the attribute graph is far from having a locally tree-like topology, is to use the classification results yielded by MLP in place of the original attribute graph, as the external field acting on the BPGCN (i.e., adopting MLP as a pre-classification step). Tests confirm that this pre-processed version of BPGCN greatly enhances the classification accuracy on the \net{Pubmed} network, up to $81.7\%$, which is significantly better than the state-of-the-art results yielded by APPNP. This pre-processing step does not improve the results on \net{Citeseer} and \net{Cora}, as expected, since the attribute graphs are sparse.

\begin{table}[h] 
  \centering
  \begin{tabular}{l|c|c|c|c|c}
       &\net{Karate}& \net{Polblogs} & \net{Citeseer}     & \net{Cora}  &\net{Pubmed} \\
  \hline
  \# nodes &34 & 1490&3327 &2078 & 19717\\
  \hline
  \# features &0 &0 &3703 &1433 &500 \\
  \hline
  \# groups &2 &2 &6 &7 &3\\
  \hline
  \# training &4 &20 &120 &140 &60\\
  \hline
  \# validation &4 &20 &500 &500 &500\\
  \hline
  \# test &26 &1450 &1000 &1000 &1000\\
  \hline
  $c$&4.6 &22.4 &2.78 &3.89 & 4.49\\
  \hline
    MLP~\cite{kipf2017semi} &--- & --- &58.4&52.2&72.7\\
    \hline
    GCN~\cite{kipf2017semi} &$\mathbf{96.5}$ &86.2 &71.1&81.5&79.0\\
    \hline
    GAT~\cite{2018graph}   &87.9 &88.7 & 70.8&83.1&78.5\\
    \hline
    SGCN~\cite{pmlr-v97-wu19e} & 91.6& 81.3&71.3&81.7&78.9\\
    \hline
    APPNP~\cite{klicpera2018combining}& 96.3 &87.5 & $\mathbf{71.8}$& $\mathbf{83.5}$& $\mathbf{80.1}$\\
    \hline
    BPGCN~    &95.8 & $\mathbf{89.3}$ &71.1& 82.1& 70.0\\
  \end{tabular}
   \vspace{0.24cm} 
  \caption{Classification performance of BPGCN and reference GCNs on real-world networks. $c$ denotes the average degree of the connectivity graph. For reference GCNs, results on \net{Karate Club} and \net{Polical blogs} are carried out using publically available implementations of these algorithms; results on \net{Cora}, \net{Pubmed} and \net{Citeseer} are adapted from ~\cite{Fey/Lenssen/2019}. For BPGCN, the reported accuracy values are based on the vanilla version. The pre-processed version of the BPGCN greatly enhances the classification accuracy on the \net{Pubmed} network, up to $81.7\%$(see text). \label{tab:results}} 
 \end{table} 
 
%

\section{Conclusions}\label{sec:con}
In this study, we constructed the joint stochastic block model (JSBM) that generates a two-fold graph which simultaneously models item nodes and feature nodes in an aggregated network setting, based on the celebrated SBM. Utilizing the cavity method in statistical physics and the corresponding belief propagation (BP) algorithm which is asymptotically exact in the thermodynamic limit, theoretical results on the detectability phase transition point and the phase diagram for JSBM are uncovered. Expectantly, JSBM could be used to generate benchmark networks with continuously tunable parameters, which might be particularly useful in evaluating the classification performance of graph neural networks.

Based on the BP equations established on JSBM, we proposed an algorithm for semi-supervised classification, adopting the graph convolution network structure, which we termed as BPGCN. In contrast to most existing graph convolution networks, the convolution kernel and activation function of BPGCN are determined mathematically from the Bayesian inference on the JSBM. We show that on synthetic networks generated by JSBM, BPGCN clearly outperforms several well-known existing graph convolution networks, and obtains extraordinary classification accuracy in the parameter regime where conventional GCNs fail to work; on real-world networks, BP also displays comparable performance to state-of-the-art GCNs. Compared with conventional GCNs, BPGCN is quite powerful in extracting label information from edge connectivities; this advantage is rooted in its non-backtracking convolution kernel inherited from the BP algorithm. The weakness of BPGCN is demonstrated by the Pubmed dataset, in which case there are too few features for the attribute graph to be approximated by a random bipartite graph; a remedy for applying BPGCN on such graphs is proposed, which uses MLP as a pre-processing step, and the corresponding advanced version of BPGCN is demonstrated. Based on the fact that BPGCN is immune to the sparsity issue and the overfitting issue exposed by conventional GCNs, we discussed possible ways inspired by BPGCN to improve current GCN techniques. It would be interesting to combine successful features of BPGCN and state-of-the-art GCNs in greater depth and design new architectures for graph neural networks; we leave this idea for future work.

A \textit{pytorch} and \textit{C++} implementation of our BPGCN and other GCNs on JSBM together with the real world datasets used in our experiments, are available at~\cite{code}.

\bibliography{ref.bbl}
\appendix
\section{Belief propagation equations}
Combine the Boltzmann distribution Eq.~\eqref{eq:bayes}, the likelihood function Eq.~\eqref{eq:likelihood}, and the variational distribution Eq.~\eqref{eq:bethe}, by minimizing the Bethe free energy with respect to constraints subject to the normalizations of marginals, one arrives at the standard form of belief propagation equations ~\cite{yedidia2003understanding} for JSBM:
\begin{widetext}
\begin{align}
\label{eq:sbp}
\psi _{t_i}^{i\rightarrow j}&=\frac{\alpha_{t_i}}{Z^{i\rightarrow j}}\prod _{k\in \partial i\setminus j}\sum _{t_k}  p_{t_it_k}\psi _{t_k}^{k\rightarrow i} \prod_{k \notin \partial i\setminus j}\sum _{t_k}(1-p_{t_it_k}) \psi_{t_k}^{k\rightarrow i}.\prod _{\mu\in\partial   i}\sum_{t_{\mu}}
q_{t_it_{\mu}}\psi_{t_{\mu}}^{\mu\rightarrow i} \prod _{\mu\notin\partial   i}\sum_{t_{\mu}}
(1-q_{t_it_{\mu}})\psi_{t_{\mu}}^{\mu\rightarrow i}\notag\\
\psi _{t_i}^{i\rightarrow {\mu}}&=\frac{\alpha_{t_i}}{Z^{i\rightarrow \mu}}\prod _{k\in \partial i}\sum _{t_k}  p_{t_it_k}\psi _{t_k}^{k\rightarrow i} \prod _{k\notin \partial i}\sum _{t_k}  (1-p_{t_it_k})\psi _{t_k}^{k\rightarrow i}\cdot \prod _{\nu\in\partial   i\setminus \mu}\sum_{t_{\nu}}
q_{t_it_{\nu}}\psi_{t_{\nu}}^{\nu\rightarrow i} \prod _{\nu\notin\partial   i\setminus \mu}\sum_{t_{\nu}}(1-q_{t_it_{\nu}})\psi_{t_{\nu}}^{\nu\rightarrow i}\notag\\
\psi_{t_{\mu}}^{\mu\rightarrow i}&=\frac{\beta_{t_{\mu}}}{Z^{\mu\rightarrow i}}   \prod _{j\notin\partial   \mu\setminus i}\sum_{t_j}
q_{t_{\mu}t_j}\psi_{t_j}^{j\rightarrow \mu}\prod _{j\in\partial   \mu\setminus i}\sum_{t_j}
(1-q_{t_{\mu}t_j})\psi_{t_j}^{j\rightarrow \mu}.
\end{align}
\end{widetext}
where $\psi _{t_i}^{i\rightarrow j}$, $\psi _{t_i}^{i\rightarrow \mu}$, and $\psi_{t_{\mu}}^{\mu\rightarrow i}$ are cavity probabilities. Marginal probabilities are estimated as a function of cavity probabilities as
\begin{align}
\label{eq:smp}
\psi _{t_i}^i&=\frac{\alpha_{t_i}}{Z^{i\rightarrow j}}\prod _{k\in \partial i}\sum _{t_k}  p_{t_it_k}\psi _{t_k}^{k\rightarrow i} \prod_{k \notin \partial i}\sum _{t_k}(1-p_{t_it_k}) \psi_{t_k}^{k\rightarrow i}\notag\\ & .\prod _{\mu\in\partial   i}\sum_{t_{\mu}}
q_{t_it_{\mu}}\psi_{t_{\mu}}^{\mu\rightarrow i} \prod _{\mu\notin\partial   i}\sum_{t_{\mu}}
(1-q_{t_it_{\mu}})\psi_{t_{\mu}}^{\mu\rightarrow i}\notag\\
\psi_{t_{\mu}}^{\mu}&=\frac{\beta_{t_{\mu}}}{Z^{\mu\rightarrow i}}   \prod _{j\notin\partial   \mu}\sum_{t_j}
q_{t_{\mu}t_j}\psi_{t_j}^{j\rightarrow \mu}\prod _{j\in\partial   \mu}\sum_{t_j}
(1-q_{t_{\mu}t_j})\psi_{t_j}^{j\rightarrow \mu}.
\end{align}
Since we have nonzero interactions between every pair of nodes, in Eq.~\eqref{eq:sbp} we have in total $n(n-1)+2mn$ messages. This results to an algorithm where even a single update takes $O(n^2)$ time, making it suitable only for networks of up to a few thousand nodes. Fortunately, for large
sparse networks, i.e., when $n$, $m$ is large and $p_{t_it_j}=q_{t_it_{\mu}}=O(1/n)$, we can neglect terms of sub-leading order in the equations. In this case,
it is assumed that $i$ or $\mu$ sends the same message to all its non-neighbors $j$, and these messages are viewed as representing an external field. By this means, now we only need to keep track of $2M$ messages, where $M$ is the total number of all edges and hyperedges, and each update step takes $O(n+m)$ time.

Suppose that $j \notin\partial i$, we have :
\begin{align}
\psi _{t_i}^{i\rightarrow j}&=\frac{\alpha_{t_i}}{Z^{i\rightarrow j}}\prod _{k\in \partial i}\sum _{t_k}  p_{t_it_k}\psi _{t_k}^{k\rightarrow i} \prod_{k \notin \partial i\setminus j}(1-\sum _{t_k}p_{t_it_k} \psi_{t_k}^{k\rightarrow i})\notag\\ & .\prod _{\mu\in\partial   i}\sum_{t_{\mu}}
q_{t_it_{\mu}}\psi_{t_{\mu}}^{\mu\rightarrow i} \prod _{\mu\notin\partial   i}\sum_{t_{\mu}}
(1-q_{t_it_{\mu}})\psi_{t_{\mu}}^{\mu\rightarrow i}\notag\\
&=\psi_{t_i}^i+O(1/n)
\end{align}
Similarly, we have $\psi_{t_{\mu}}^{\mu\rightarrow i}=\psi_{t_{\mu}}^{\mu}+O(1/m)$. To the leading order, the messages on non-edges do not depend on the target node.
For nodes with $j \in\partial i$, we have
\begin{align}
\psi _{t_i}^{i\rightarrow j}&=\frac{\alpha_{t_i}}{Z^{i\rightarrow j}}\prod _{k\in \partial i\setminus j}\sum _{t_k}  p_{t_it_k}\psi _{t_k}^{k\rightarrow i} \prod_{k \notin \partial i}(1-\sum _{t_k}p_{t_it_k} \psi_{t_k}^{k\rightarrow i})\notag\\ & .\prod _{\mu\in\partial   i}\sum_{t_{\mu}}
q_{t_it_{\mu}}\psi_{t_{\mu}}^{\mu\rightarrow i} \prod _{\mu\notin\partial   i}(1-\sum_{t_{\mu}}
q_{t_it_{\mu}}\psi_{t_{\mu}}^{\mu\rightarrow i})\notag\\
&\simeq \alpha_{t_i}\frac{e^{-h_{t_i}}}{Z^{i\rightarrow j}}\prod _{k\in \partial i\setminus j}\sum _{t_k}  p_{t_it_k}\psi _{t_k}^{k\rightarrow i}   \prod _{\mu\in\partial   i}\sum_{t_{\mu}}
q_{t_it_{\mu}}\psi_{t_{\mu}}^{\mu\rightarrow i}
\end{align}
where terms having $O(1/n)$ and $O(1/m)$ contribution to $\psi_{t_i}^{i\rightarrow j}$ combining to the definition of the auxiliary external field:
\begin{align}
h_{t_i}&=\sum _k\sum _{t_k}p_{t_i t_k}\psi _{t_k}^k+\sum_{\mu} \sum_{t_{\mu}}q_{t_i t_{\mu}}\psi_{t_{\mu}}^{\mu}
\end{align} 
Applying the same approximations to the external fields acting on feature nodes, one finally arrives at BP equations~\eqref{eq:bp}.

\section{Detectability transition analysis using noise perturbations}
On a graph generated by the JSBM with $n$ item nodes and $m$ feature nodes, the probability that an item node has $k_1$ neighboring item nodes $p_1(k_1)$ follows a Poisson distribution with average degree $c_1$, and the probability of it being associated with $k_2$ feature nodes $p_2(k_2)$ follows a Poisson distribution with average degree $c_2$. Similarly, the probability that a feature node has $k_3$ neighboring item nodes $p_3(k_3)$ follows a Poisson distribution with average degree $c_3 = nc_2/m$. 
Considering a branching process on a graph generated by JSBM with infinite size, the average branching ratio of the process is related to the excess degree which is defined upon the average number of neighbors. The average excess degree of an item node is computed as
\begin{align}
\widetilde{c}_1=\frac{\sum_{k_1}k_1(k_1-1)p_1(k_1)}{\sum_{k_1}k_1p_1(k_1)}=c_1.
\end{align}
Similarly we have $\widetilde{c}_2=c_2$ as well, and the excess degree of feature nodes is 
\begin{align}
\widetilde{c}_3=\frac{\sum_{k_2}p_2(k_2)k_2(k_2-1)n/m}{\sum_{k_2}k_2p_2(k_2)}=c_2n/m=c_3.
\end{align}

Consider the noise propagation process on a tree graph as depicted in  Fig.~\ref{fig:noise_pp}, with the number of depth $l\to\infty$. In the tree, odd layers contain exclusively item nodes, and even layers contain feature nodes (blue boxes) and edges (green boxes) which are connected to item nodes in odd layers. Assume that on the leaves of the tree (nodes on the $l$-th layer) the paramagnetic fixed point is perturbed as
\begin{align}
\psi_{t_{v_l}}^{v_l\rightarrow v_{l-1}}=\alpha_{t_{v_l}}+\delta_{t_{v_l}}^{v_l\rightarrow v_{l-1}}.
\end{align}
where $v_l$ represent an item node on the $l$ layer, $t_{v_l}$ is label of $v_l$, $v_0$ corresponds to the root node $i$ in Fig.~\ref{fig:noise_pp}. Now let us investigate the influence of the perturbation, from the message on any leaf, to the message on the root node. For simplicity, first choose one path only containing item nodes (i.e., not connected via any feature node) and latter generalize to paths containing both item nodes and feature nodes. We define the Jacobian matrix $\mathbf{T}^{i\to j}$ with respect to the message passing from an item node $i$ to another item node $j$ along an edge $(ij)$, and the Jacobian matrix $\mathbf{T}^{\mu\to i}$ corresponding to the message passing from a feature node $\mu$ to an item node $i$, and similarliy the matrix $\mathbf{T}^{i\to \mu}$ for the backward passage. Elements of these Jacobian matrices are formulated as 
\begin{align}
T^{i\rightarrow j}_{t_it_k}&=\frac{\partial \psi _{t_j}^{j\rightarrow k}}{\partial\psi _{t_i}^{i\rightarrow j}} \big|_{\alpha_{t_i}}=\alpha_{t_i}(\frac{np_{t_i t_k}}{c_1}-1),\notag\\
T^{\mu\rightarrow i}_{t_{i}t_{\mu}}&=\frac{\partial \psi _{t_i}^{i\rightarrow j}}{\partial\psi _{t_{\mu}}^{\mu\rightarrow i}} \big|_{\alpha_{t_{i}},\beta_{t_{\mu}}}=\alpha_{t_i}(\frac{m q_{t_{\mu} t_i}}{c_2}-1),\notag\\
T^{i\rightarrow \mu}_{t_it_{\mu}}&=\frac{\partial \psi _{t_{\mu}}^{\mu\rightarrow j}}{\partial\psi _{t_i}^{i\rightarrow \mu}} \big|_{\alpha_{t_i},\beta_{t_{\mu}}}=\beta_{t_{\mu}}(\frac{m q_{t_i t_{\mu}}}{c_2}-1).
\end{align}
These matrices represent propagation strength~\eqref{eq:bp} between two messages in the vicinity of the paramagnetic fixed point. We can see that, all three matrices are independent of node indices, only depend on the type of the nodes. If the path only contains edges, the perturbation $\delta^{v_0\rightarrow x}_{t_{v_0}}$ on the root node induced by the perturbation $\delta^{v_l \rightarrow v_{l-1} }_{t_{v_l}}$ on the leaf node can be written as 
\begin{align}
  \delta^{v_0\rightarrow x}_{t_{v_0}}=\sum_{\{t_{v_d}:d=1,...,l\}}\prod_{d=0}^{l-1} T^{i\rightarrow j}_{t_{v_d},t_{v_{d+1}}} \delta^{v_l \rightarrow v_{l-1}}_{t_{v_l}},
\end{align}
or in the vector form, $\mathbf \delta^{v_0 \rightarrow x}=(\mathbf T^{i\rightarrow j})^l\mathbf \delta^{v_l\rightarrow v_{l-1}}$. Now consider the path contains both edges and hyper edges: every time a hyper edge is passed through, $\mathbf T^{i\rightarrow \mu}$ and $\mathbf T^{\mu\rightarrow i}$ transmits to $\mathbf T^{i\rightarrow j}$; therefore, the total weight acting on the path is 
$$\mathbf \delta^{v_0\rightarrow x}=(\mathbf T^{i\rightarrow j})^s(\mathbf T^{i\rightarrow \mu}\mathbf T^{\mu\rightarrow i})^{(l-s)}\mathbf\delta^{v_l\rightarrow v_{l-1}}$$ 
where $s$ is the number of edges and $l-s$ is the number of feature nodes on this path. 

For $l\to\infty$, $(\mathbf T^{i\rightarrow j})^s(\mathbf T^{i\rightarrow \mu}\mathbf T^{\mu \rightarrow i})^{(l-s)}$ is dominated by the product of the largest eigenvalues, $\lambda_A$, $\lambda_F$, and $\lambda_{F'}$ of the three matrices; notice that $\lambda_F=\lambda_{F'}$. So the above equation can be written as $$\mathbf\delta^{v_0\rightarrow x}=(\lambda_A)^s(\lambda_F)^{2(l-s)}\delta^{v_l\rightarrow v_{l-1}}.$$ 
Then consider the collection of all perturbations on the root node $v_0$ from all leaves. Obviously the mean is $0$, and the variance on the $t$-th component of the perturbation vector is computed as
\begin{align}
    &\left \langle(\delta^{v_0\rightarrow x}_t)^2\right\rangle\nonumber\\
    &=\left\langle \left(\sum_{s=0}^{l}\sum_{1}^{\binom{l}{s}}\sum_{v_l=0}^{(c1)^s(c2c_3)^{l-s}}{\lambda_A}^s{\lambda_F}^{2(l-s)}\delta^{v_l\rightarrow v_{l-1}}\right)^2\right\rangle\notag\\
    &=\sum_{s=0}^{l}\sum_{1}^{\binom{l}{s}}\sum_{v_l=0}^{(c1)^s(c2c_3)^{l-s}}{\lambda_A}^{2s}{\lambda_F}^{4(l-s)}\left \langle(\delta_t^{v_l\rightarrow v_{l-1}})^2\right\rangle\notag\\
    &=\sum_{s=0}^{l}\binom{l}{s}(c_1{\lambda_A}^2)^s(c_2c_3{\lambda_F}^4)^{l-s}\left \langle(\delta_t^{v_l\rightarrow v_{l-1}})^2\right\rangle\notag\\
    &=(c_1{\lambda_A}^2+c_2c_3{\lambda_F}^4)^{l}\left \langle(\delta_t^{v_l\rightarrow v_{l-1}})^2\right\rangle.
\end{align}
Here, we have made use of the property that all perturbations on different leaves are independent. Obviously, the paramagnetic fixed point is locally unstable under random perturbations when $(c_1{\lambda_A}^2+c_2c_3{\lambda_F}^4)^{l}>1$, so the phase transition of detectablity locates at
\begin{align}
c_1{\lambda_A}^2+c_2c_3{\lambda_F}^4 = 1. 
\end{align}

%

\section{Graph convolution networks and the spectral localization problem of linear convolution kernels}
Convolution networks~\cite{lecun1995convolutional} have been proved to be one of the most successful models for image classifications~\cite{krizhevsky2012imagenet} and many other machine learning problems~\cite{goodfellow2016deep}. The success of CNN is credited to the convolution kernel which is a linear operator defined on grid-like structures. However, in recent years, a significant amount of attention has been paid to generalizing convolutional operations to graphs, which do not have grid structures.

In \cite{bruna2013spectral}, authors propose to define convolutional layers on graphs that operate on the spectrum of the graph Laplacian:
\begin{equation}
\mathbf X_{l+1}=\sigma(\Lambda \star \mathbf X_l)=\sigma\left(  \mathbf V \mathbf \Lambda \mathbf V^{T}\mathbf X_{l}     \right),
\end{equation}
where $\mathbf X_{l}$ is the state in the $l^{\textrm{th}}$ layer, $\mathbf V$ is the eigenvector matrix of the graph Laplacian $\mathbf L=\mathbf I-\mathbf D^{-\frac{1}{2}}\mathbf A\mathbf D^{-\frac{1}{2}}$ ($\mathbf A$ is adjacency matrix and $\mathbf D$ is the diagonal matrix on node degrees), and $\sigma(\cdot)$ is an element-wise non-linear activation function, and $\mathbf \Lambda$ is a diagonal matrix representing a kernel in the frequency (graph Fourier) domain. Usually $\mathbf \Lambda$ contains only several non-zero elements, working as cut-offs to the frequency in the Fourier domain, because it is believed that using only a few eigenvectors of the graph Laplacian are sufficient for describing smooth structures of the graph. Only computing leading eigenvectors is also computationally efficient.

Even though, computing only a few eigenvectors of the Laplacian of large graphs could be quite cumbersome. In \cite{hammond2011wavelets}, authors proposed to parametrize the kernel in the frequency domain through a truncated series expansion, using the Chebyshev polynomials up to $K$-th order:
\begin{equation}
\mathbf X_{l+1}=\sigma(\Lambda \star \mathbf X_l)\approx \sigma\left[ \sum_{k=0}^Kc_kT_k\left(\frac{2}{\lambda_{\textrm{max}}}\mathbf L-\mathbf I \right)\mathbf X_{l}\right],
\end{equation}
where $c_k$ denotes coefficients, $T_K(x)=2xT_{k-1}(x)-T_{k-2}(x)$ are recursively defined Chebyshev polynomials, and $\lambda_{max}$ is the largest eigenvalue.
In~\cite{kipf2017semi} the authors limited the convolution operation to $K=1$, and approximate 
$\lambda_{\textrm{max}}=2$,
then obtain
$$\mathbf X_{l+1}=\sigma(\mathbf \Lambda \star \mathbf X_l)\approx 
\theta_1X-\theta_2D^{-\frac{1}{2}}\mathbf A\mathbf D^{-\frac{1}{2}},$$
with two free parameters $\theta_1$ and $\theta_2$.
Further, \cite{kipf2017semi} restricted the two parameters by specifying 
$\theta=\theta_1=-\theta_2$,
and introduced a normalization trick 
$$\mathbf I+\mathbf D^{-\frac{1}{2}}\mathbf A\mathbf D^{-\frac{1}{2}}\rightarrow  \widehat {\mathbf D}^{-\frac{1}{2}}\widehat {\mathbf A}\widehat {\mathbf D}^{-\frac{1}{2}},$$
where $\mathbf {\widehat A}=\mathbf A+\mathbf I$, and $\widehat D_i=1+\sum_j A_{ij}$. Finally, upon generalizing to node features of $\kappa$ components, that is, to the $\kappa$-channel signal $\mathbf X\in\mathbb{R}^{n\times \kappa}$,
one arrives at the GCN~\cite{kipf2017semi}, with the forward model taking form of
\begin{equation}\label{eq:gcn1}
\mathbf X_{l+1}=\sigma(\widetilde {\mathbf A} \mathbf X_{l}\mathbf W),
\end{equation}
where $\sigma(\cdot)$ is an activation function such as ReLU, $\mathbf X_{l}\in \mathbb{R}^{n\times \kappa_{l}}$ is the network state at layer $l$ ($n$ denotes the number of nodes in the graph, $\kappa_{l}$ is the dimension of state at layer $l$). Matrix $\mathbf W\in\mathbb{R}^{\kappa_{l+1},\kappa_l}$ is the trainable weight matrix at the $l$-th layer, and the propagator $\widetilde {\mathbf A}$ ~\eqref{eq:A} is responsible for convolving the neighborhood of a node, which isshared over the whole network.

It has been shown in~\cite{kipf2017semi} that GCN significantly outperforms related standard methods, including manifold regularization (ManiReg)~\cite{weston2012deep}, Semi-supervised embedding (SemiEmb)~\cite{yang2016revisiting}, label propagation (LP)~\cite{zhu2003semi}, skip-gram based graph embedding (DeepWalk)~\cite{perozzi2014deepwalk}, the iterative classification algorithm in conjunction with two classifiers taking care of both local node features and aggregations~\cite{lu2003link}, and the recently proposed Planetoid~\cite{yang2016revisiting}.

For many further developments of GCNs in recent years, almost all of them can be understood as an effective object representation starting from the original feature vector and then projected forward by multiplying finite times of the linear convolution kernel. So it is recognized that the main principle of graph convolution kernels is inspired by the spectral properties of linear operators, such as Laplacians, normalized Laplacians~\cite{kipf2017semi,pmlr-v97-wu19e}, random walk matrix~\cite{klicpera2018combining} etc. The underlying assumption is that the eigevectors of the graph convolution kernel contain global information about group labels, which can be revealed during the forward propagation of GCNs. 

However, it is known that this assumption may not hold on large sparse networks, because the eigenvectors of conventional linear operators such as graph Laplacians are subject to the localization problem, induced by the fluctuation of degrees or the local structures of the graph~\cite{krzakala2013spectral,zhang2016robust}. Even on random graphs where the Poisson degree distribution is rather concentrated, the localization problem is still significant on large graphs. For the adjacency matrix, it is well known that the largest eigenvalue is bounded below by the squared root of the largest node degree (which grows as $\frac{\log(n)}{\log\log(n)}$), and diverges on large graphs when the number of nodes $n\to\infty$. Thus the corresponding eigenvectors only report information of the largest degrees. For normalized matrices such as the normalized Laplacian, there are many eigenvalues that are very close to $0$, with corresponding eigenvectors reporting information about local dangling sub-graphs rather than global structures related to the group labels. 
We refer to \cite{krzakala2013spectral,zhang2016robust} for detailed analysis of spectrum localizations of graph Laplacians, and for the comparison between spectral algorithms using different operators on large sparse graphs. Unfortunately, real-world networks are usually sparse, as we can see from Table~\ref{tab:results} that (although they might not be large enough) the citation networks we used for experiments all have an average degree around $4$, which is extremely low. 

On large synthetic networks the sparsity issue for conventional GCNs is revealed more clearly (see main text). We carried out extensive experiments to verify our analysis (in Fig.~\ref{fig:more}). 
 
\section{Belief Propagation Graph Convolution Network(BPGCN)}
On a graph generated by JSBM with $n$ item nodes, $m$ features nodes, $nc_1/2$ edges and $nc_2$ hyper edges, i.e. the average degree of item nodes in the connectivity graph is $c_1$ and the average degree of the attribute graph is $c_2$. We define matrices storing cavity messages $\mathbf \Psi^{i\rightarrow j}_l \in \mathbb R^{nc_1\times \kappa}$, $\mathbf \Psi^{i\rightarrow \mu }_l\in \mathbb R^{nc_2\times \kappa}$ and $\mathbf \Psi^{\mu\rightarrow i }_l\in \mathbb R^{nc_2\times \kappa}$, and marginal matrices $\mathbf \Psi^i_l\in \mathbb R^{n\times \kappa}$ and $\mathbf \Psi^{\mu}_l\in \mathbb R^{m \times \kappa}$, where $l$ indicates the cavity messages after the $l$-step of iterations, The BP equations~\eqref{eq:bp} can be written in the form of matrix multiplications:
\begin{align}
   \label{eq:mbp}
&\mathbf \Psi_{l+1}^{\mu}=\mathrm{Softmax}\left[ \mathbf{U_I}\log(\mathbf\Psi^{i\to \mu}_{l}\mathbf P)\right]\notag\\
      &\mathbf\Psi^i_{l+1}=\mathrm{Softmax}\left[ \mathbf U_{II}\log(\mathbf\Psi^{I \rightarrow i }_l\mathbf Q)+\mathbf U_{III}\log(\mathbf\Psi^{i \rightarrow j}_l\mathbf P)\right]\notag\\
    &\mathbf \Psi^{i\rightarrow j}_{l+1}=\mathrm{Softmax}\left[\mathbf B_I\text{log}(\mathbf\Psi^{\mu \rightarrow i }_l\mathbf Q)+\mathbf B_{II}\log(\mathbf \Psi^{i \rightarrow j}_l\mathbf P)\right]\notag\\
    &\mathbf\Psi^{i\rightarrow \mu}_{l+1}=\mathrm{Softmax}\left[ \mathbf B_{III}\log(\mathbf \Psi_l^{i \rightarrow j}\mathbf P)+\mathbf B_{IV}\log(\mathbf \Psi^{\mu\rightarrow i}_l \mathbf Q)\right]\notag\\
    &\mathbf\Psi^{\mu\rightarrow i}_{l+1}=\mathrm{Softmax}\left[\mathbf B_{V}\log(\mathbf\Psi^{i \rightarrow \mu }\mathbf Q)\right]
\end{align}
where $\mathbf P$, $\mathbf Q$ are affinity matrices of size $\kappa \times \kappa$.

Kernel matrices $\mathbf{U_I}$ to $\mathbf{U_{III}}$, $\mathbf{B_I}$ to $\mathbf{B_V}$ are non-backtracking matrices~\cite{krzakala2013spectral} that encode adjacency information of cavity messages and marginals, which are defined as:
\begin{align}
\label{eq:nb}
&\mathbf U_I^{\mu,i\to \nu}=\delta_{\mu \nu} \in\{0,1\}^{m\times nc_2}\notag\\
&\mathbf U_{II}^{i,\mu\to j}=\delta_{ij} \in\{0,1\}^{n\times nc_2}\notag\\
&\mathbf U_{III}^{i,k\to l}=\delta_{il}\in\{0,1\}^{n\times nc_1}\notag\\
&\mathbf B_I^{i\to j,\mu\to l}=\delta_{il}(1-\delta_{\mu j})\in\{0,1\}^{nc_1\times nc_2}\notag\\
&\mathbf B_{II}^{i\to j,k\to l}=\delta_{il}(1-\delta_{kj})\in\{0,1\}^{nc_1\times nc_1}\notag\\
&\mathbf B_{III}^{i\to \mu,j\to l}=\delta_{il}(1-\delta_{\mu j})\in\{0,1\}^{nc_2\times nc_1}\notag\\
&\mathbf B_{IV}^{i\to \mu,\nu\to l}=\delta_{il}(1-\delta_{\mu \nu})\in\{0,1\}^{nc_2\times nc_2}\notag\\
&\mathbf B_{V}^{\mu\to i,j\to \nu}=\delta_{\mu \nu}(1-\delta_{ij})\in\{0,1\}^{nc_2\times nc_2}.
\end{align}
Therefore, the matrix multiplication form of the parallel-updated BP equations can be viewed as a forward model of a neural network. When the above equations propagate, and are truncated after $L$ steps of iterations, the resulting algorithm scheme is termed as BPGCN, with $L$ layers. States in the last layer are extracted as the output of BPGCN for computing the cross-entropy loss on training labels ~\eqref{eq:loss}. Matrices $\mathbf Q$ and $\mathbf P$ are trainable parameters of the BPGCN and are updated using the back propagation algorihtm. Input of the neural network are probability-normalized random initial cavities and marginals. In order to accelerate the training process of BPGCN, we use an adjustable external field strength $\gamma$ as a hyper parameter. For example, suppose node $i$ is in the training set, i.e. we have label $t_i$ for node $i$, a term $\gamma\log(0.1,0.1,0.9,....1)$ is added to \eqref{eq:mbp} inside the $\text{Softmax}$ function. When $\gamma$ approaches infinity, node $i$ is pinned to label $t_i$ \cite{Zhang2014phase}, and is used as an hyper parameter adjusted by the validation set. For nodes in the training set, cavity probabilities and marginals are initialized as $(0,1,...,0)$. Thus the parameters of BPGCN are $\mathbf P$ and $\mathbf Q$, and the hyper parameters are $\epsilon_1$, $\epsilon_2$, and $\gamma$. The hyper parameters we used in the experiments on real-world datasets (Table.~\ref{tab:results} are:
for \net{Citeseer} and \net{Cora}, $L=5$ and external strength $\gamma=2.0$; on \net{Cora} we set $\epsilon_1=0.1$, $\epsilon_2=0.6$; on \net{Citeseer} we set $\epsilon_1=0.1$, $\epsilon_2=0.5$. For \net{Pubmed}, $L=2$, external strength $\gamma=1.5$, $\epsilon_1=0.3$, and $\epsilon_2=0.8$; when the attribute graph is pre-processed with a MLP to offer BPGCN as a prior or an external field, we set $L=12$, external strength $\gamma=0.5$, $\epsilon_1=0.1$.   For both \net{Karate Club} and \net{Political Blogs} networks, $L=5$, external strength $\gamma=0.5$, and $\epsilon_1=0.1$ ($\epsilon_2$ is not applicable).

\begin{figure}
    \centering
    \includegraphics[width=0.41\textwidth]{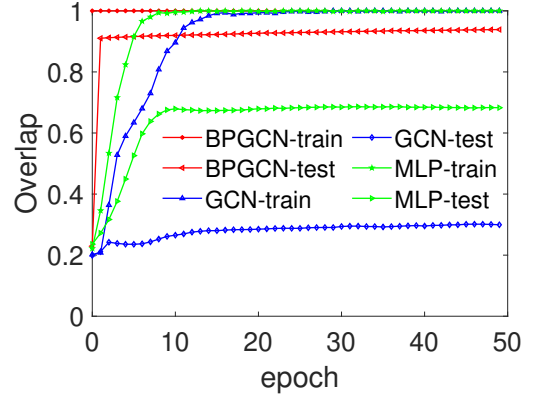}
    \caption{Training and test overlap (Eq.~\eqref{eq:ovl}) as a function of epoch on a synthetic graph generated by JSBM, with $n=m=10000$, $c_1=c_2=10$, $\epsilon1=1$, and $\epsilon_2=0.1$.}
    \label{fig:train_process}
\end{figure}

\section{More comparisons on synthetic networks}
\begin{figure*}[h]
\centering
\includegraphics[width=0.31\textwidth]{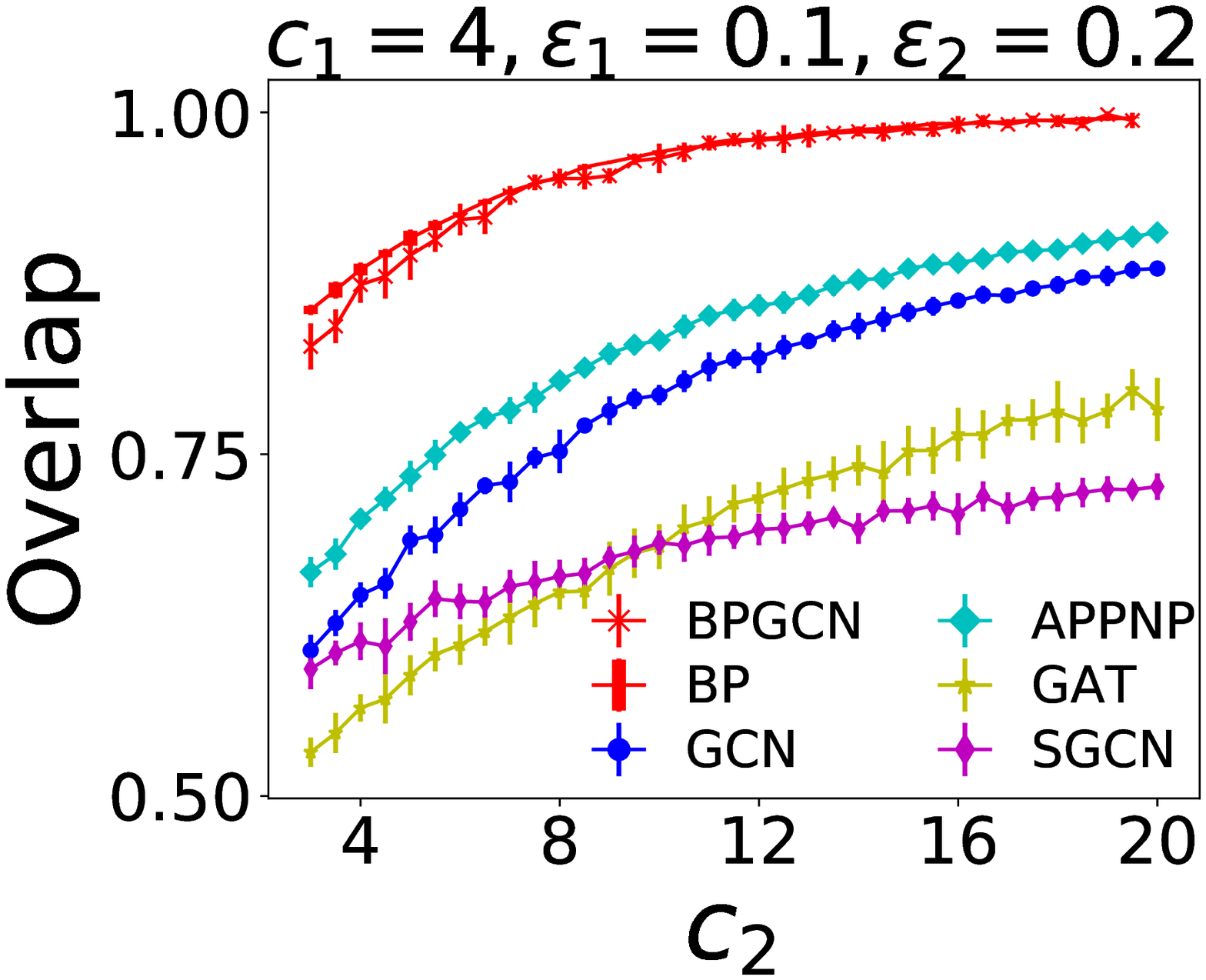}
\includegraphics[width=0.31\textwidth]{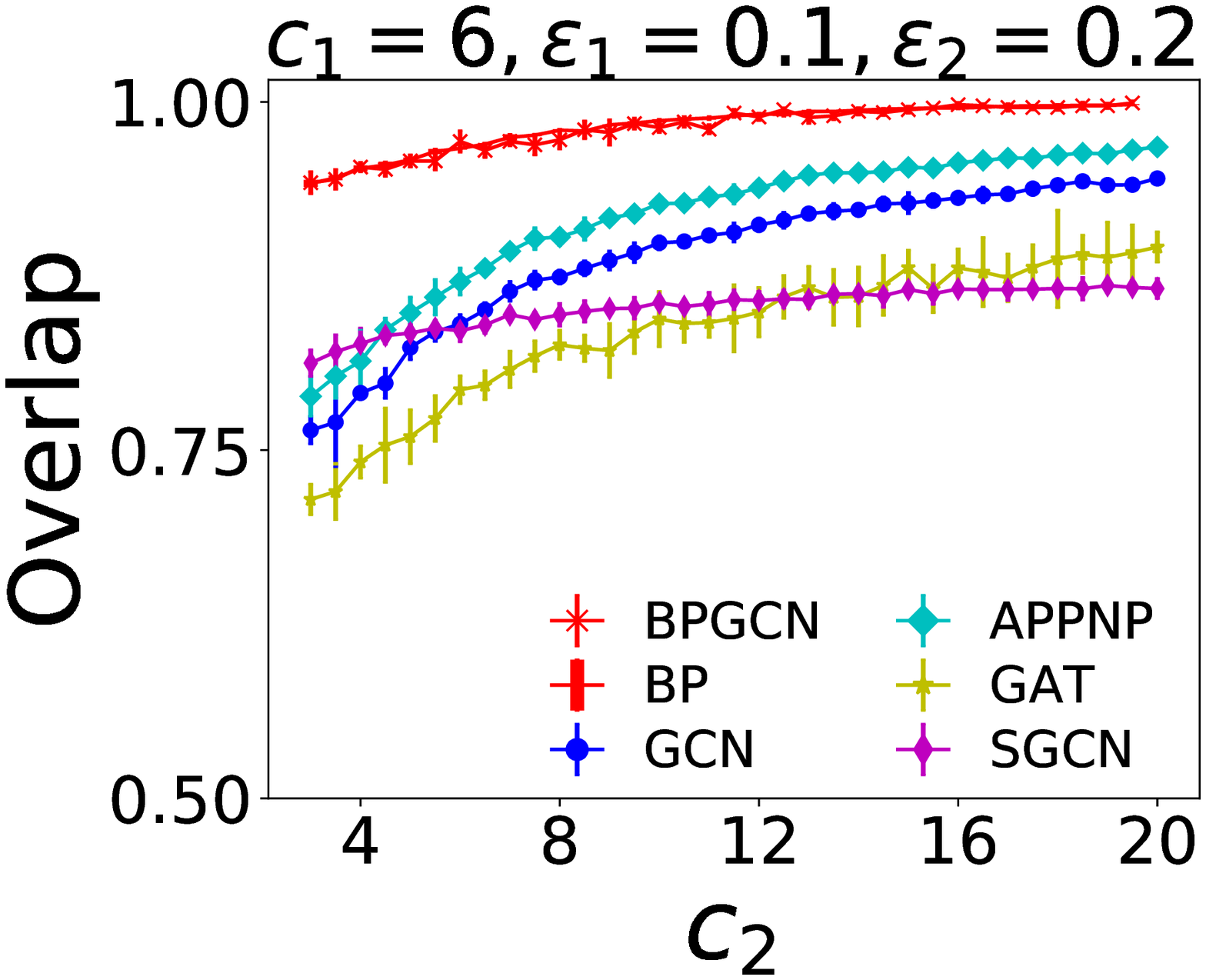}
\includegraphics[width=0.31\textwidth]{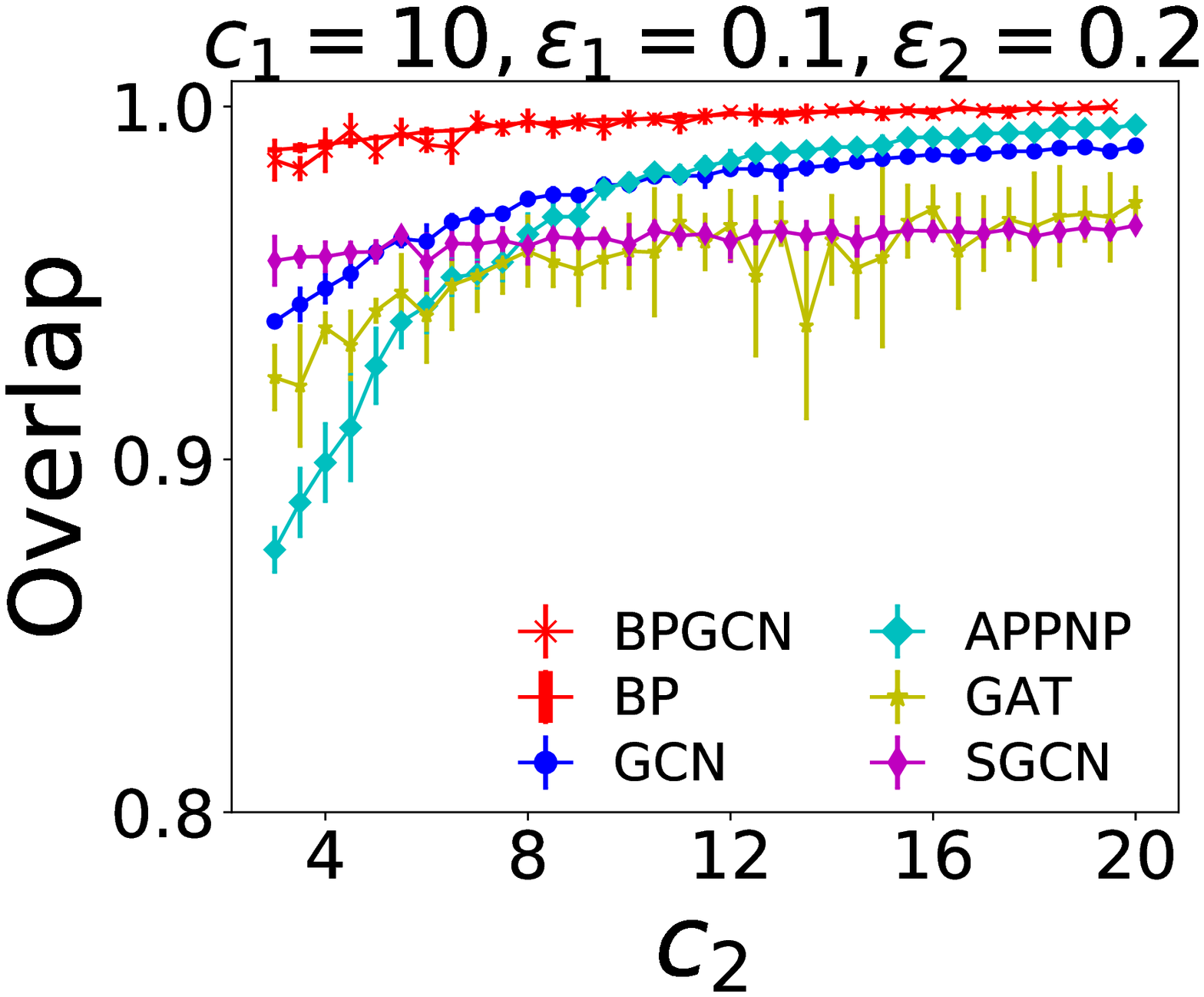}
\includegraphics[width=0.31\textwidth]{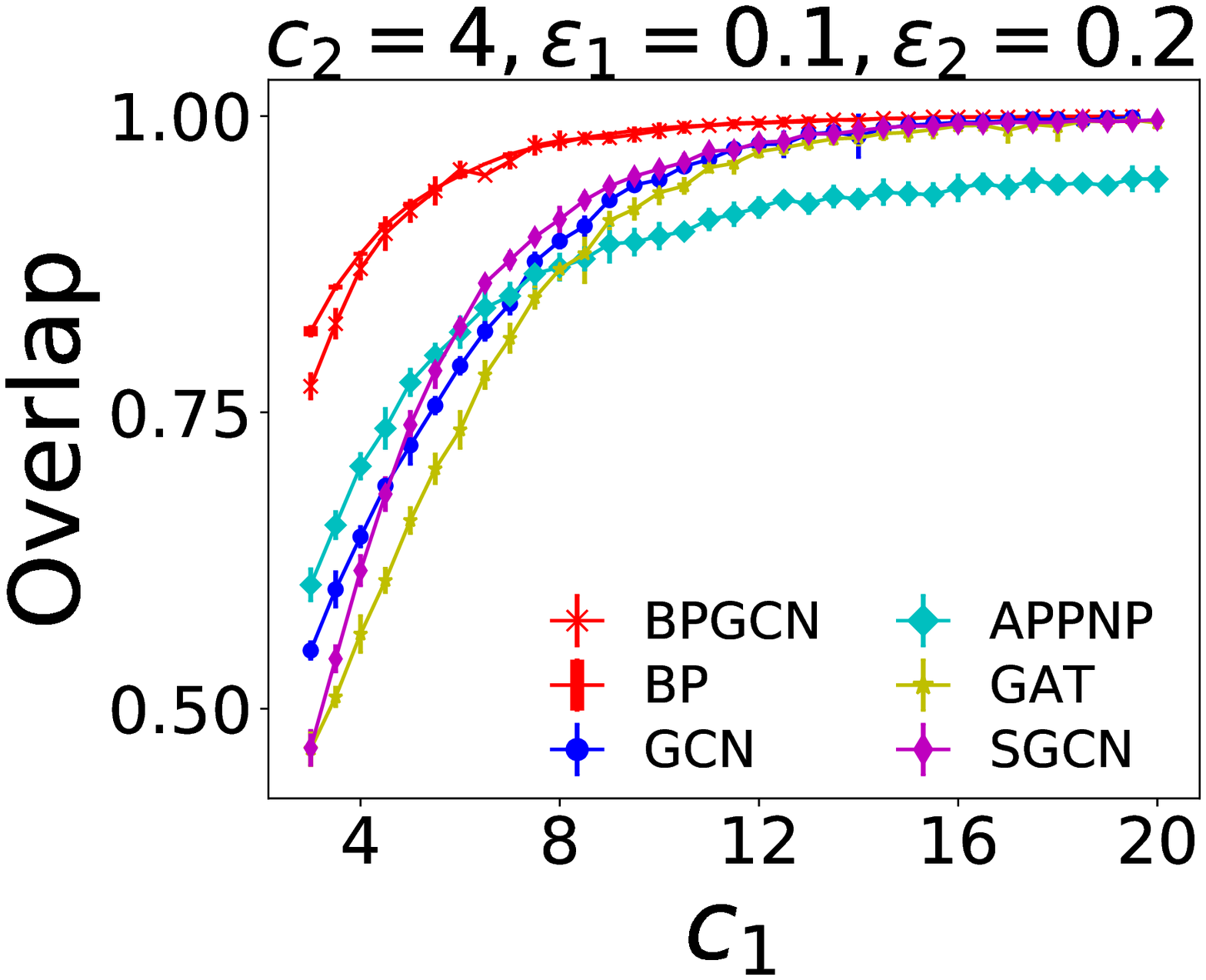}
\includegraphics[width=0.31\textwidth]{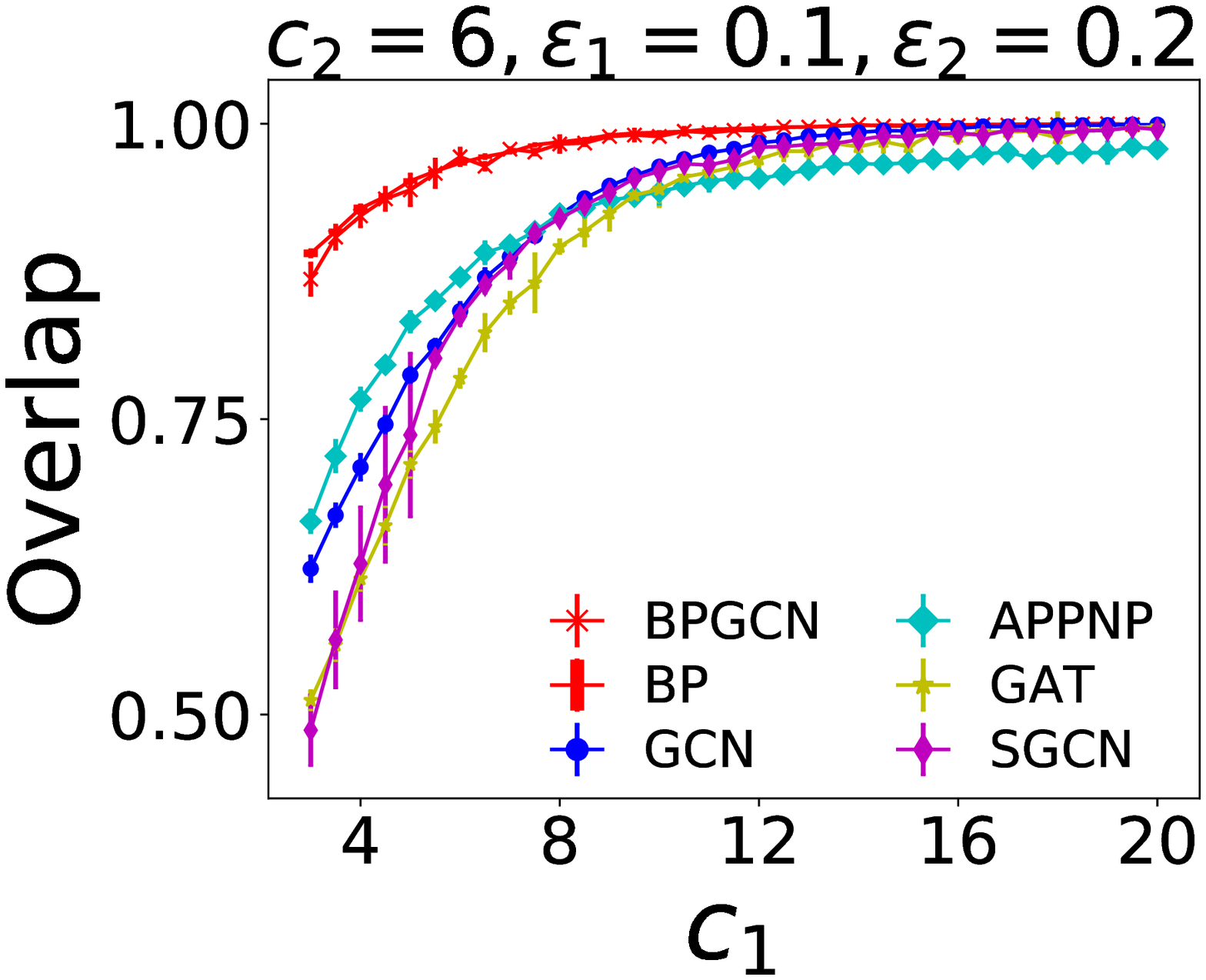}
\includegraphics[width=0.31\textwidth]{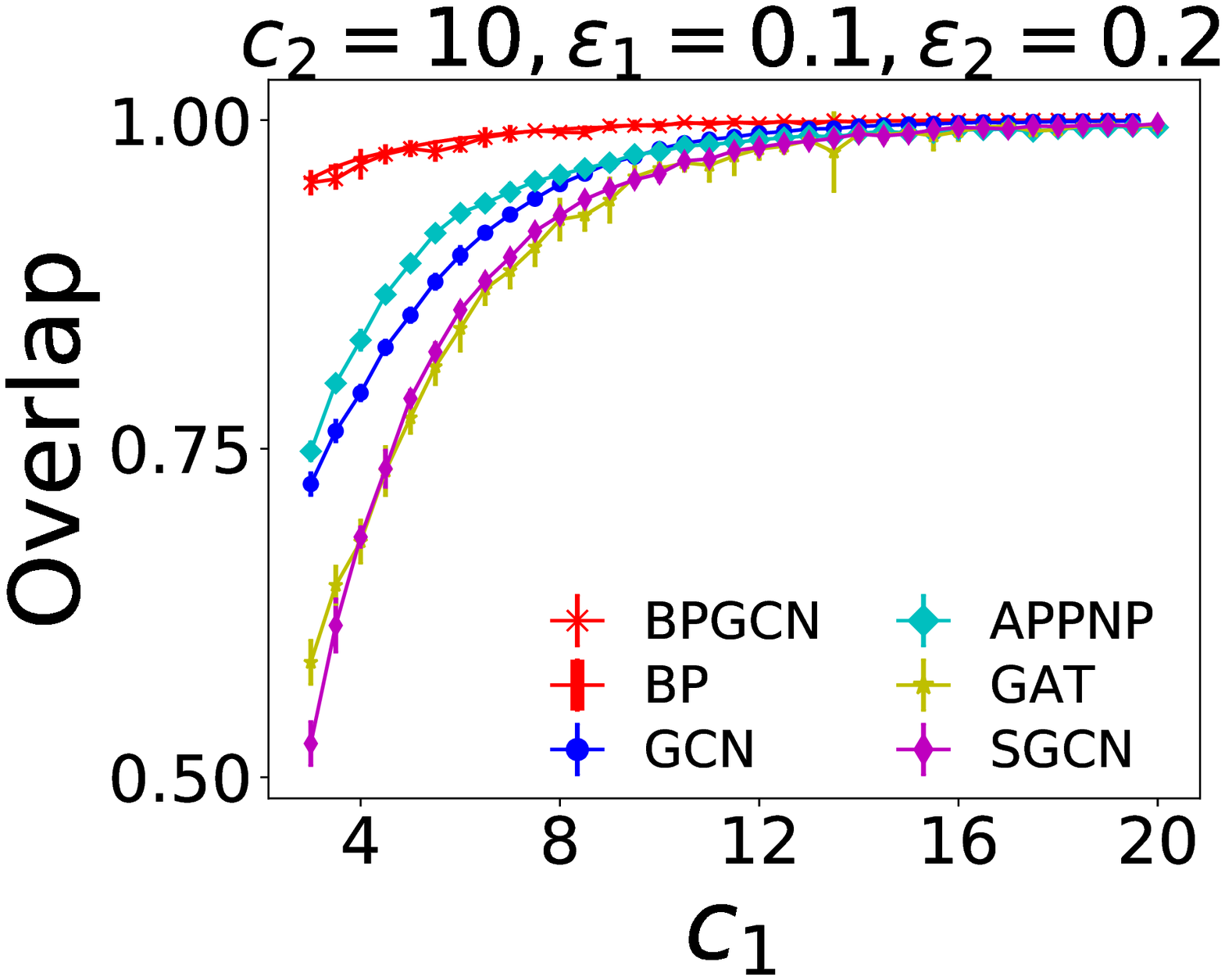}
\includegraphics[width=0.31\textwidth]{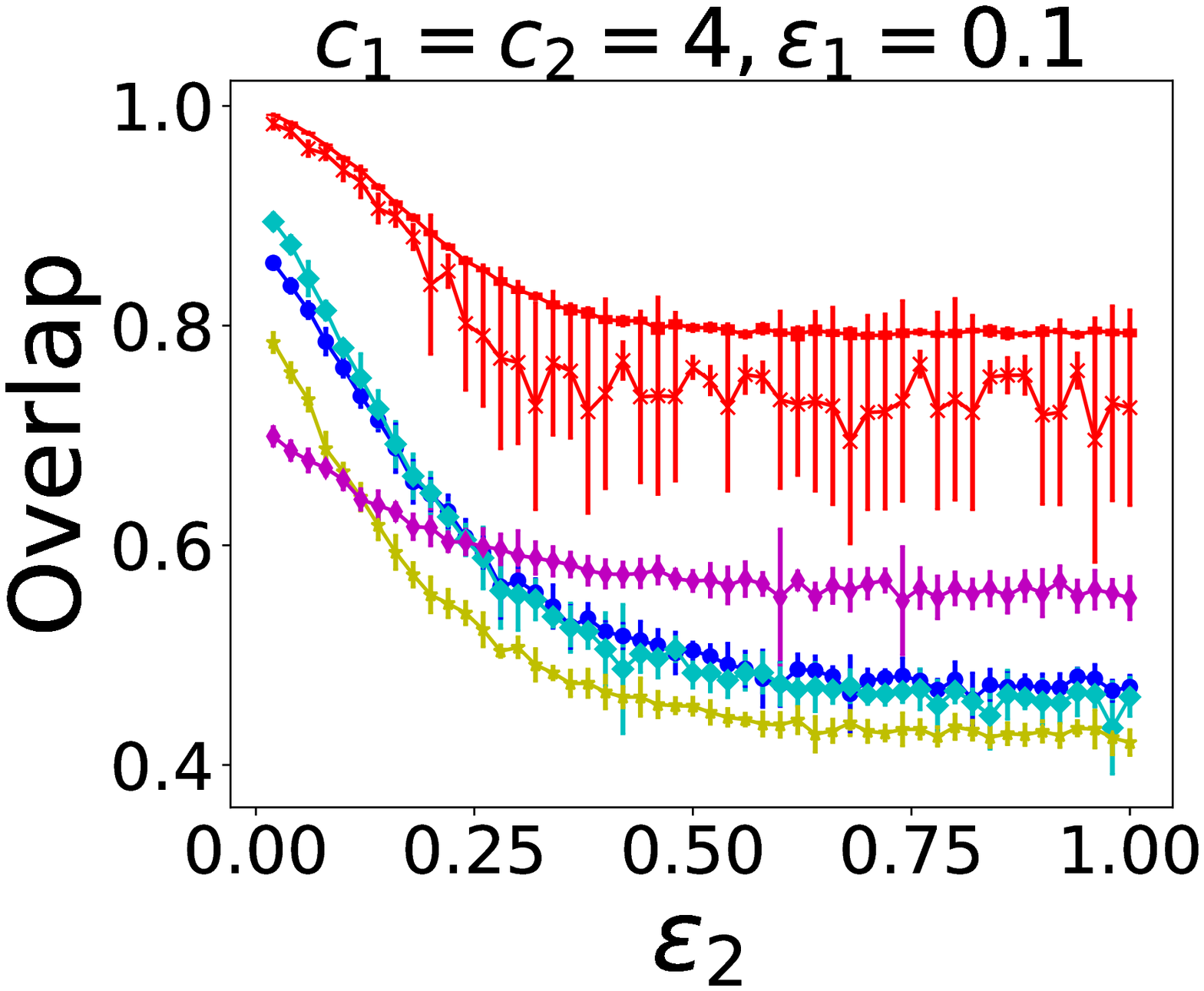}
\includegraphics[width=0.31\textwidth]{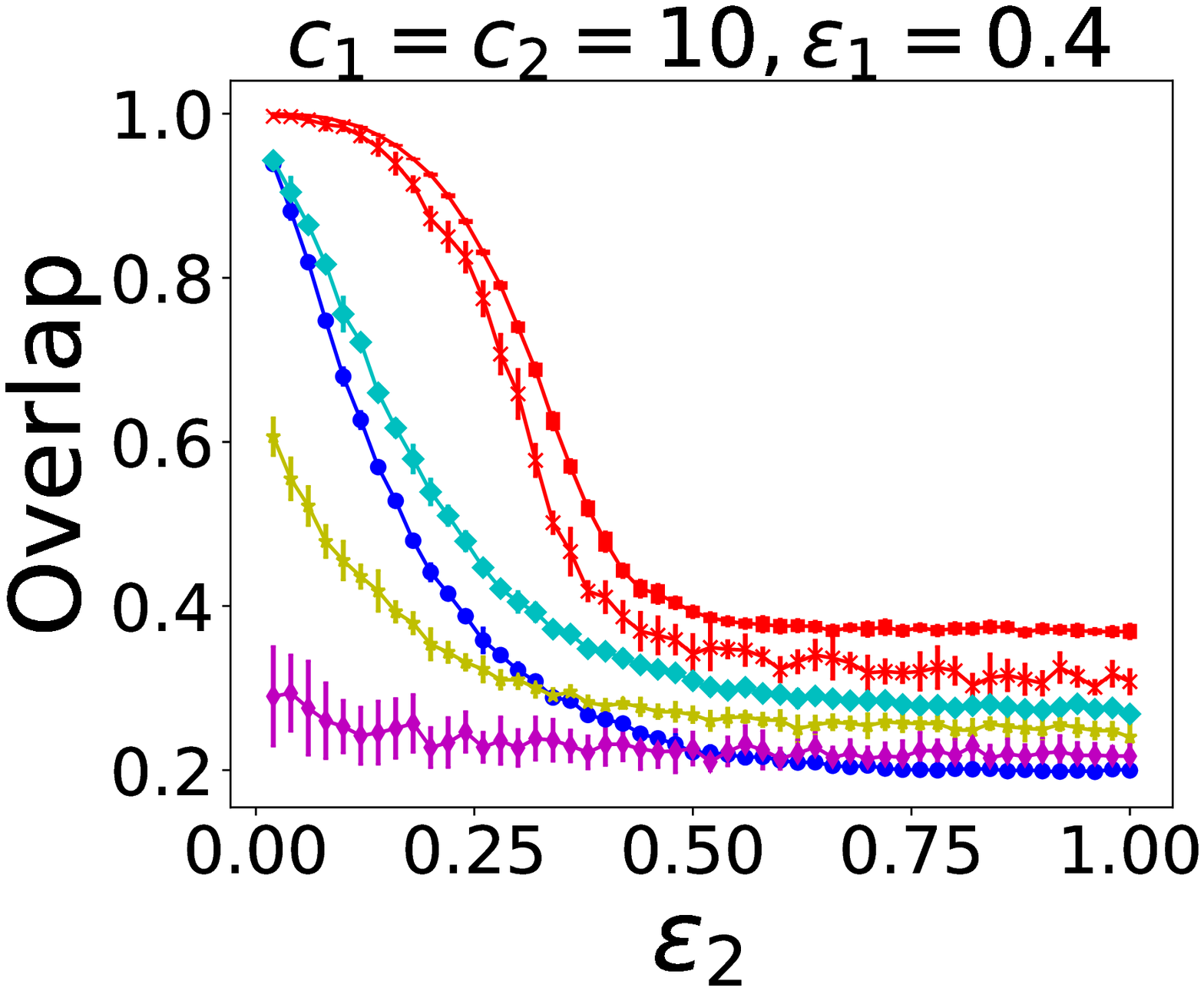}
\includegraphics[width=0.31\textwidth]{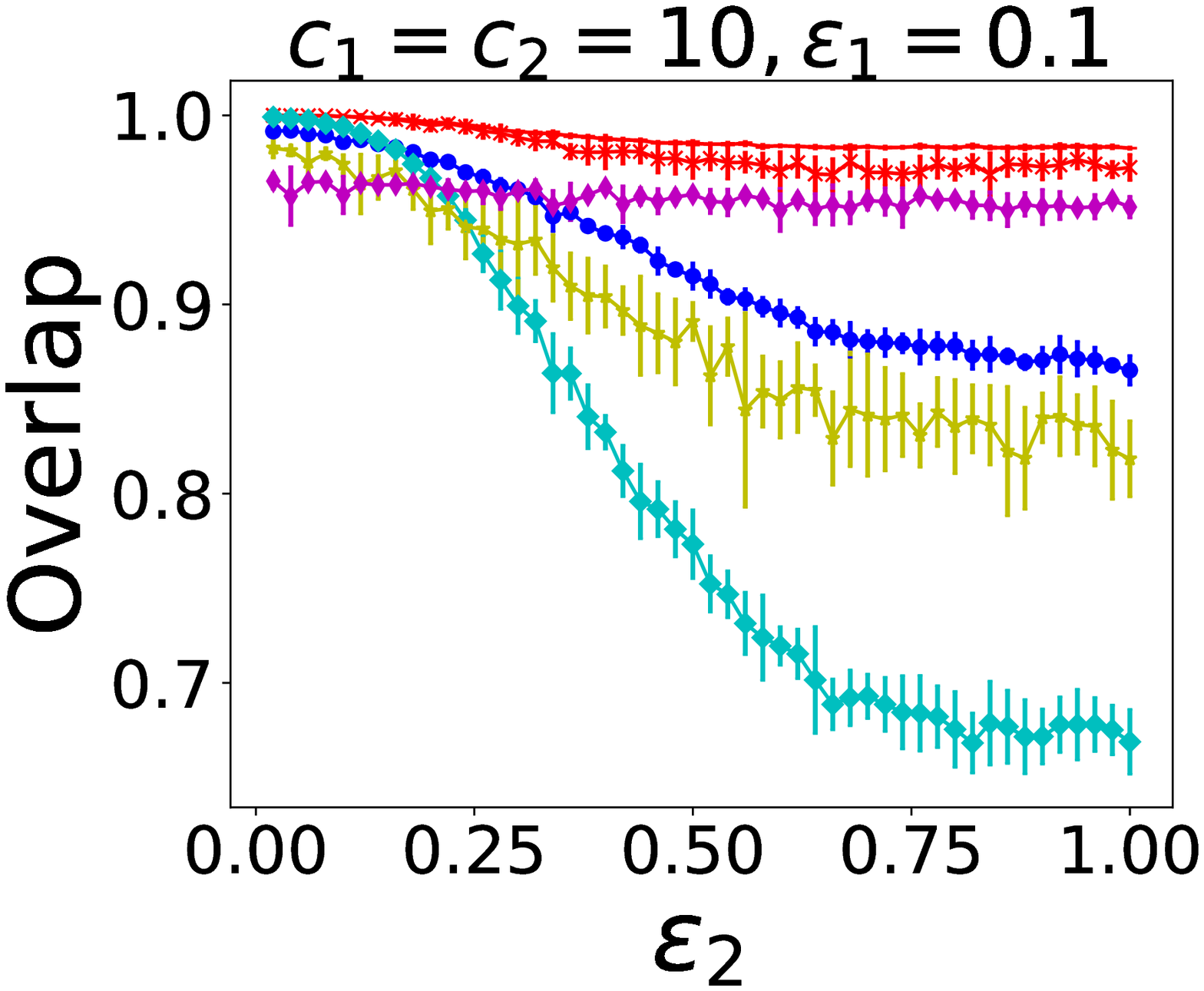}
\includegraphics[width=0.31\textwidth]{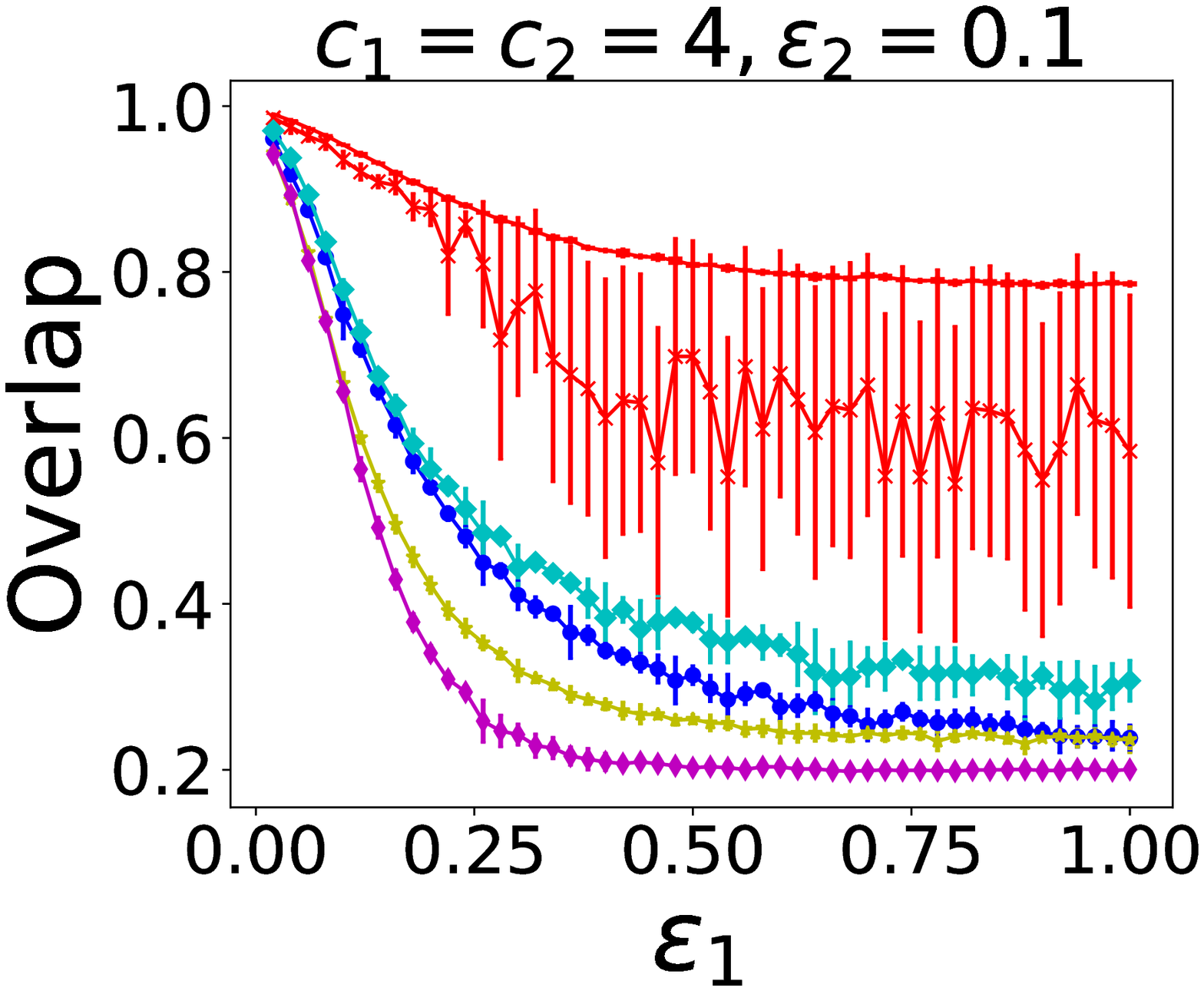}
\includegraphics[width=0.31\textwidth]{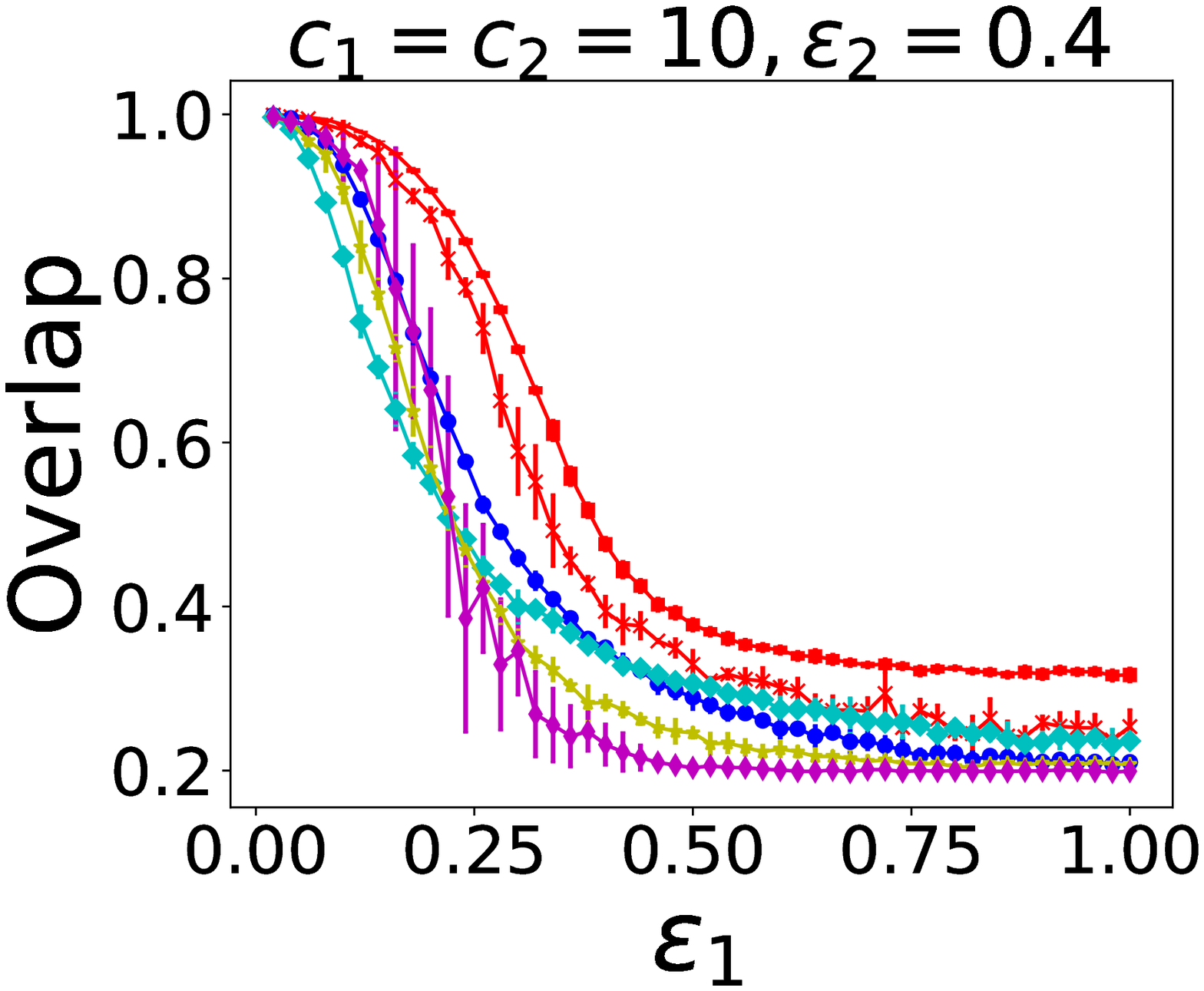}
\includegraphics[width=0.31\textwidth]{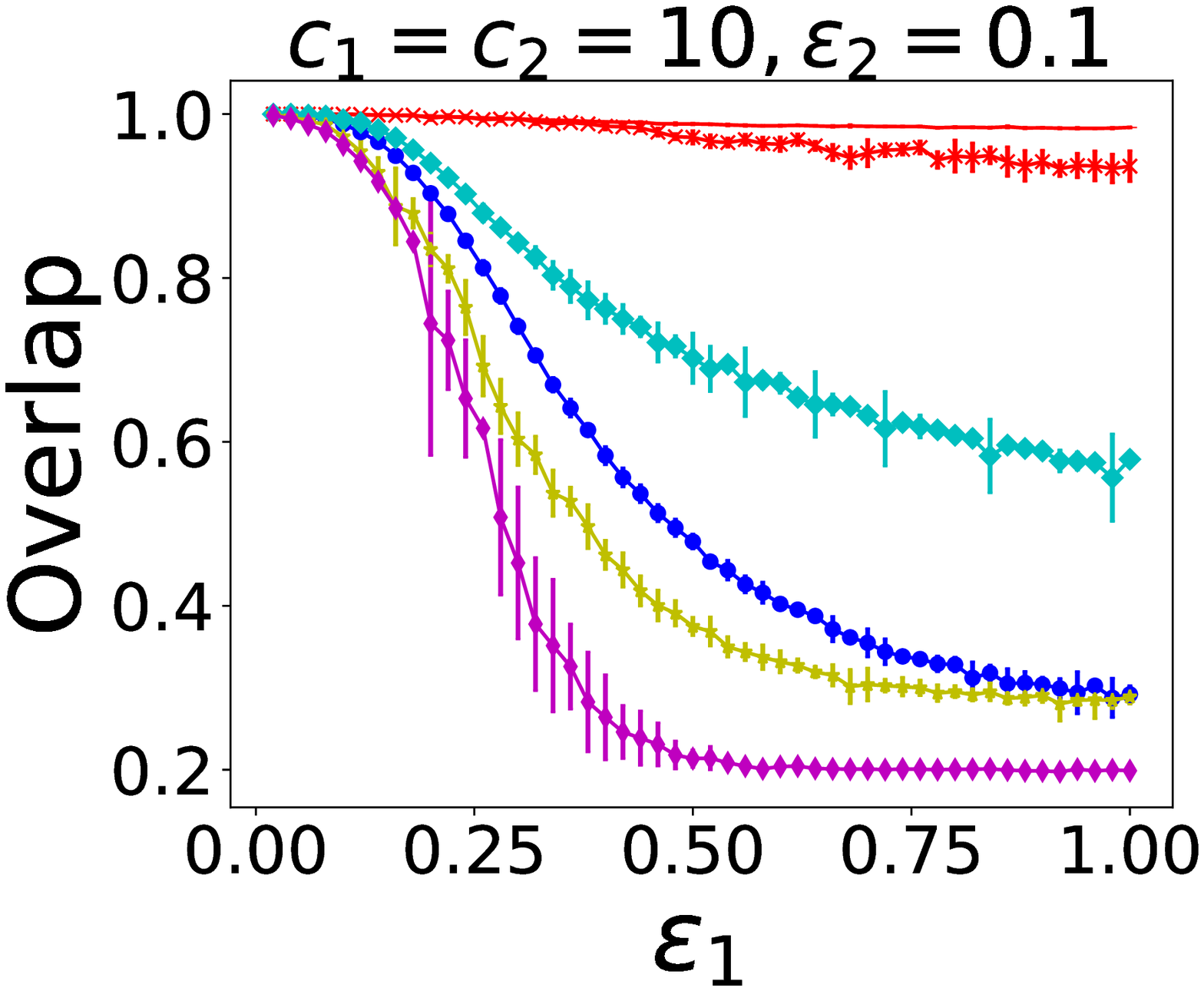}
\caption{Overlap (Eq.~\eqref{eq:ovl}) obtained by BP, BPGCN and reference GCNs on synthetic networks generated by JSBM with different parameters. All networks have $n=m=10000$, and group numbers $\kappa=5$. The fraction of training labels is fixed at $\rho=0.05$. Each data point is averaged over $10$ random instances. The other parameters are printed above each subfigure.\label{fig:more}} 
\end{figure*}

We carried out extensive numerical experiments on synthetic networks generated by JSBM with various parameters. The overlap of BP, BPGCN and reference algorithms BPGCN, GCN, APPNP, GAT and SGCN are compared in Fig.\ref{fig:more}. In the top row, the average degree of the connectivity graph is fixed to $c_1=4, 6$, and $10$ from left to right. Due to the spectrum localization problem of graph Laplacians, as we have discussed earlier, when $c_1$ is fixed to $4$, GCN, APPNP, GAT, and SGCN work much worse than BPGCN, even when $c_2$ is large. We also see that when $c_1$ is large, most of these tested GCNs work well, even for a very low $c_2$. In the second row, the average degree of the attribute graph $c_2$ is fixed and $c_1$ varies. Performances of reference GCNs approaches the performance of BPGCN, when $c_1$ gradually increases. 

In the third row of the figure, $c_1, c_2$ and $\epsilon_1$ are fixed and $\epsilon_2$ varies. On the left, $c_1=c_2=4, \epsilon_1=0.1$. The graphs are in the sparse regime so conventional GCNs do not work well due to the sparsity issue. In the middle, $c_1=c_2=10, \epsilon_1=0.4$. Graphs are not sparse, but the edges contains relatively little information about group labels. Results show that SGCN has trouble in this regime even when $\epsilon_2$ is very small. On the right $c_1=c_2=10$, and $\epsilon_1=0.1$. The graphs are not in the sparse regime, and edges contain enough information about labels. We see that all GCNs except APPNP perform reasonably well; APPNP has trouble with low $\epsilon_2$, probably because an non-optimal $\omega$ parameter was learned.

In the last row of the figure, $c_1$, $c_2$, and $\epsilon_2$ are fixed, and the inverse signal-to-noise ratio $\epsilon_1$ varies. On the left, $c_1=c_2=4, \epsilon_2=0.1$. We see that BPGCN starts to have a large variance, but the overlap is on average much higher than other GCNs, due to the sparsity of the graphs. In the middle, $c_1=c_2=10$, $\epsilon_2=0.4$, meaning that the hyper edges in the attribute graph contain little information of labels, and the major information comes from the connectivity graph which is not sparse. In this case we see that the performances of reference GCNs are comparable to BP and BPGCN, since there is no sparsity issue. On the right, $c_1=c_2=10$, and $\epsilon_2=0.1$. There is no sparsity issue for conventional GCNs, and the attribute graph contains enough information about the labels. However, the result is quite surprising: while BP and BPGCN gives almost $100\%$ classification accuracy in the full range of $\epsilon_1$, conventional GCNs yield very low accuracy with $\epsilon_1$ larger than $0.5$. This phenomenon implies that, when the edges of the graph are quite noisy, information from the features is not well-extracted by conventional GCNs. 

To understand the reason for this observation, in Fig.~\ref{fig:train_process} we plot the training process of GCN, MLP and BPGCN, on a graph generated with $c_1=c_2=10$, $\epsilon_1=1$ and $\epsilon_2=0.1$. In this case it means that the connectivity graph is a pure random graph, with edges containing no information of the label at all ($\epsilon_1=1$), while the attribute graph contains adequit informatino about the ground-truth ($\epsilon_2=0.1$). We can see that BPGCN yields very high accuracy on the training set and the test set, even from the beginning, because $\epsilon_2$ is so small that the initial values for BPGCN are already good enough. At the begining of the training process, MLP has low training accuracy as well as low test accuracy. But after several epochs of training, its training accuracy goes to $1$, and it also generalizes very well to the test set on which the accuracy gradually increases to around $0.6$. However, this generalization does not happen for GCN, as we can see that although GCN fits very well to the training set, the overlap for the test set is only slighly above $0.2$, i.e., result of a random guess. Obviously, GCN overfits to the training set; the reason is that GCN assumes that edges always contain information of the labels, even when the underlying graph is purely random. Moreover, we can deduct directly from the formula of APPNP~\eqref{eq:appnp} that, since APPNP uses ratios $\omega$ and $1-\omega$ to weight the contributions of edges and features, the best possible result given by APPNP on graphs of the above type, where edges contain no information and features contain all the information, is almost equal to the result of MLP. Indeed, we have tested with $\epsilon_1=1, \epsilon_2=0.1, c_1=c_2=10$; APPNP achieves an overlap around $0.6$, while MLP falls around $0.66$.

\end{document}